\documentclass{emulateapj}
\usepackage[]{natbib}
\shortauthors{R. E. Louis et al.}
\pdfoutput = 1
\begin{document}

\title{Analysis of a Fragmenting Sunspot using {\em Hinode} Observations}

\author{Rohan E. Louis\altaffilmark{1}$^{\textrm{,}}$\altaffilmark{4}, B. Ravindra\altaffilmark{2}, 
Shibu K. Mathew\altaffilmark{1}, Luis R. Bellot Rubio\altaffilmark{3}, 
A. Raja Bayanna\altaffilmark{1} \\ and P. Venkatakrishnan\altaffilmark{1}}

\altaffiltext{1}{Udaipur Solar Observatory, Physical Research Laboratory
                     Dewali, Badi Road, Udaipur,
	       	     Rajasthan - 313004, India}

\altaffiltext{2}{Indian Institute of Astrophysics, 
		     II Block, Koramangla,
		     Bangalore - 560 034, India}

\altaffiltext{3}{Instituto de Astrof\'{\i}sica de Andaluc\'{\i}a (CSIC),
                     Apartado de Correos 3004,
                     18080 Granada, Spain}
\altaffiltext{4}{Presently at Leibniz-Institut f$\ddot{\textrm{u}}$r Astrophysik Potsdam (AIP),
                     An der Sternwarte 16,
                     14482 Potsdam, Germany}
\email{rlouis@aip.de}
 
\begin{abstract}
We employ high resolution filtergrams and polarimetric measurements
from {\em Hinode} to follow the evolution of a sunspot for eight days
starting on June 28, 2007.  The imaging data were corrected for
intensity gradients, projection effects, and instrumental stray light
prior to the analysis. The observations show the formation of a light
bridge at one corner of the sunspot by a slow intrusion of
neighbouring penumbral filaments. This divided the umbra into two
individual umbral cores. During the light bridge formation, there was a 
steep increase in its intensity from 0.28 to 0.7 $I_{\rm QS}$ in nearly 
4 hr, followed by a gradual increase to quiet Sun (QS) values in 13 hr.
This increase in intensity was accompanied by a large reduction in the 
field strength from 1800 G to 300 G.
The smaller umbral core gradually broke away from the parent sunspot 
nearly 2 days after the formation of the light bridge rendering the parent 
spot without a penumbra at the location of fragmentation. The penumbra in the
fragment disappeared first within 34 hr, followed by the fragment whose area
decayed exponentially with a time constant of 22 hr. In comparison, the parent
sunspot area followed a linear decay rate of 0.94 Mm$^2$~hr$^{-1}$. The depleted 
penumbra in the parent sunspot regenerated when the inclination of 
the magnetic field at the penumbra-QS boundary became within 40$^\circ$ 
from being completely horizontal and this occurred near the end of
the fragment's lifetime. After the disappearance of the fragment, another 
light bridge formed in the parent which had similar properties as the 
fragmenting one, but did not divide the sunspot. The significant 
weakening in field strength in the light bridge along with the presence 
of granulation is suggestive of strong convection in the sunspot which 
might have triggered the expulsion and fragmentation of the smaller
spot. Although the presence of QS photospheric conditions in sunspot
umbrae could be a necessary condition for fragmentation, it is not a
sufficient one.
\end{abstract}


\section{Introduction}
\label{intro} 
The solar magnetic field is distributed on a wide range of
spatial scales. Sunspots can be regarded as the largest magnetic
structures with diameters of 20-40 Mm and field strengths in excess of
2.5 kG at the photosphere (see \citet{Solanki2003a} and
\citet{Borrero2011} for reviews on the properties of sunspots). 
The average lifetime of a sunspot is of the order of several days
\citep{Ringnes1964} and its equilibrium configuration is determined 
by the balance of forces due to gravity, the magnetic field, and the
gas pressure \citep{Priest1982}.

The formation of sunspots is initiated by the coalescence of smaller
magnetic elements which rise from the convection zone to the surface
due to buoyancy.  The systematic merging of these fragments often
results in a pore which is characterized by strong and relatively
vertical fields ranging from 1 to 1.5 kG. Partial penumbral formation
occurs when the magnetic flux exceeds 1--1.5$\times 10^{20}$ Mx
\citep{Leka1998}. The penumbra develops very rapidly, with pieces of 
it being completed within an hour or less
\citep{Bumba1965,Keppens1996}.  A newly formed penumbral segment is
practically indistinguishable from a more mature one in terms of field
strengths, inclination angles and continuum intensities
\citep{Leka1998}. Recent high resolution observations by
\citet{Schlichenmaier2010} illustrate the rapid sector-wise formation
of the penumbra, which fills half the umbral circumference in about 4
hr. The development of a rudimentary penumbra initiates the Evershed
flow \citep{Evershed1909}, a nearly horizontal outflow of plasma that
starts in the inner penumbra and returns to the solar surface in the
mid penumbra and beyond. During the formation of sunspots, the umbra
is often divided by one or several light bridges which tend to
demarcate individual magnetic regions \citep{Bumba1965}. Light bridges
can also be seen during sunspot fragmentation
\citep{Bumba1965,Garcia1987}.  They are considered `field-free'
intrusions in umbrae \citep{Parker1979,Arnab1986} or manifestations of
magneto-convection
\citep{Rimmele1997,Hirzberger2002}.

According to \citet{McIntosh1981}, the decay of sunspots starts almost
as soon as they are formed. Observations by \citet{Wallenhorst1982}
indicated the absence of spreading or diffusion during the decay,
suggesting the in-situ disappearance of the magnetic field. However,
it is believed that the decay of sunspots could occur through ohmic
diffusion across a current sheet around the sunspot
\citep{Gokhale1972} or through moving magnetic features 
(MMFs)---magnetic elements that stream from the outer penumbra into
the surrounding moat. MMFs are observed as extensions of penumbral
filaments \citep{Alberto2005,Cabrera2006,Ravindra2006,Alberto2008},
sometimes driven by `Evershed clouds'
\citep{Shine1994,Cabrera2007,Cabrera2008}---velocity patches that
appear inside the penumbra and propagate radially outward along
penumbral filaments. The net flux transported out of a sunspot by MMFs
is estimated to be (0.4-6.2)$\times10^{19}$ Mx~hr$^{-1}$
\citep{Hagenaar2005}.  Thus, a sunspot of $10^{22}$ Mx would
disintegrate within one or several weeks if the sunspot flux is
removed by MMFs alone.  Recent observations of a sunspot region by
\citet{Kubo2008} indicate a flux change rate and a transport rate of
$1.2 \times 10^{19}$ Mx~hr$^{-1}$ and 2.8 $\times10^{19}$
Mx~hr$^{-1}$, respectively. Although MMFs are indicative of sunspot
decay \citep{Harvey1973}, their origin could be related to the
interaction between penumbral field lines, the Evershed mass 
flow, and the moat flow, not to the decay process itself
\citep{Valentin2002,Kubo2007}.

It is important to understand the nature and role of small-scale
instabilities in the decay of sunspots.  While detailed 3D MHD
simulations of active region formation are available 
\citep[e.g.,][]{Cheung2010}, the process of sunspot fragmentation has
not been modeled yet and is therefore poorly understood. In this
paper, the evolution of a fragmenting sunspot is investigated using
high resolution filtergrams and spectropolarimetric measurements from
the Japanese satellite {\em Hinode} \citep{Kosugi2007}. The long time
sequences provided by {\em Hinode} are ideally suited for this kind of
study. We observed and analyzed the following processes: i) the
formation of a light bridge in the umbra; ii) the fragmentation of the
sunspot at the position of the light bridge; iii) the decay of the
fragment; iv) the restoration of the penumbra in the parent sunspot;
v) the area and flux decay of the spot; vi) the rotation of the
fragment about the parent; and vii) the change in horizontal proper
motions caused by the fragmentation. The aim of this work is to
identify in-situ conditions leading to fragmentation and the different
phases of sunspot evolution so as to provide useful constraints for 3D
MHD models. Section~\ref{data} describes the data processing and
Section~\ref{res} presents the main results of the analysis. We
summarize the sunspot evolution and discuss our findings in
Section~\ref{summary}.

\begin{figure}[!h]
\centerline{
\hspace{-25pt}
\includegraphics[angle=0,width = 0.5\textwidth]{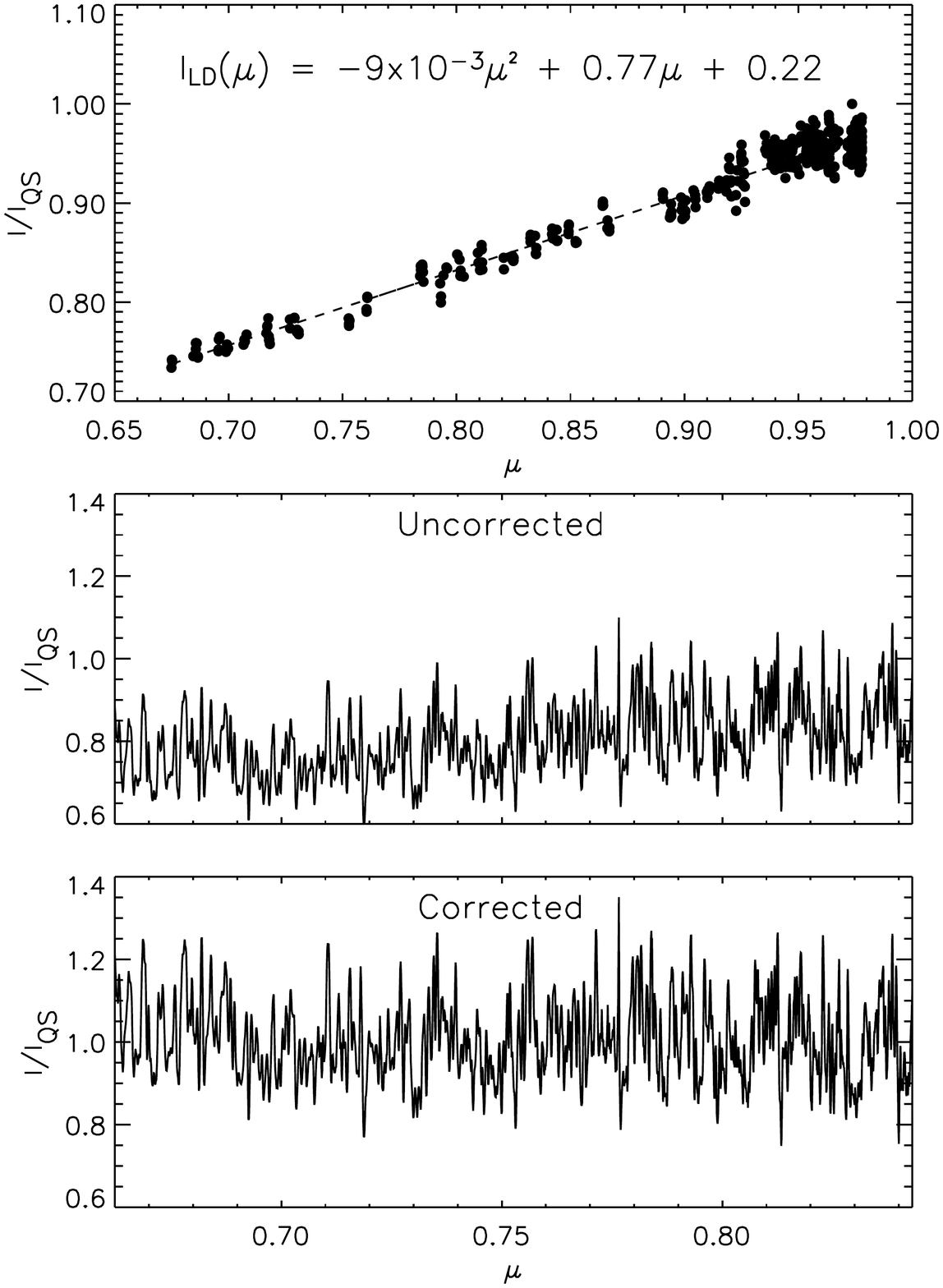}
}
\vspace{-10pt}
\caption{Limb darkening and its correction. {\em Top panel:}  
Normalized QS intensity averaged over a $100 \times 100$ pixel area in
the Level-1 G-band images ({\em black circles}) as a function of
$\mu$. The {\em dashed} line is a second order polynomial fit with
the coefficients inscribed in the plot. {\em Middle panel:} G-band
intensity along a horizontal cut in the QS, for a filtergram acquired on
June 28, 2007 at $\theta = 38^{\circ}$. {\em Bottom panel:} Same as
above, but with limb darkening removed.}
\label{limbpro}
\end{figure}

\begin{figure}[!h]
\centerline{
\hspace{-40pt}
\includegraphics[angle=0,width = 0.78\textwidth]{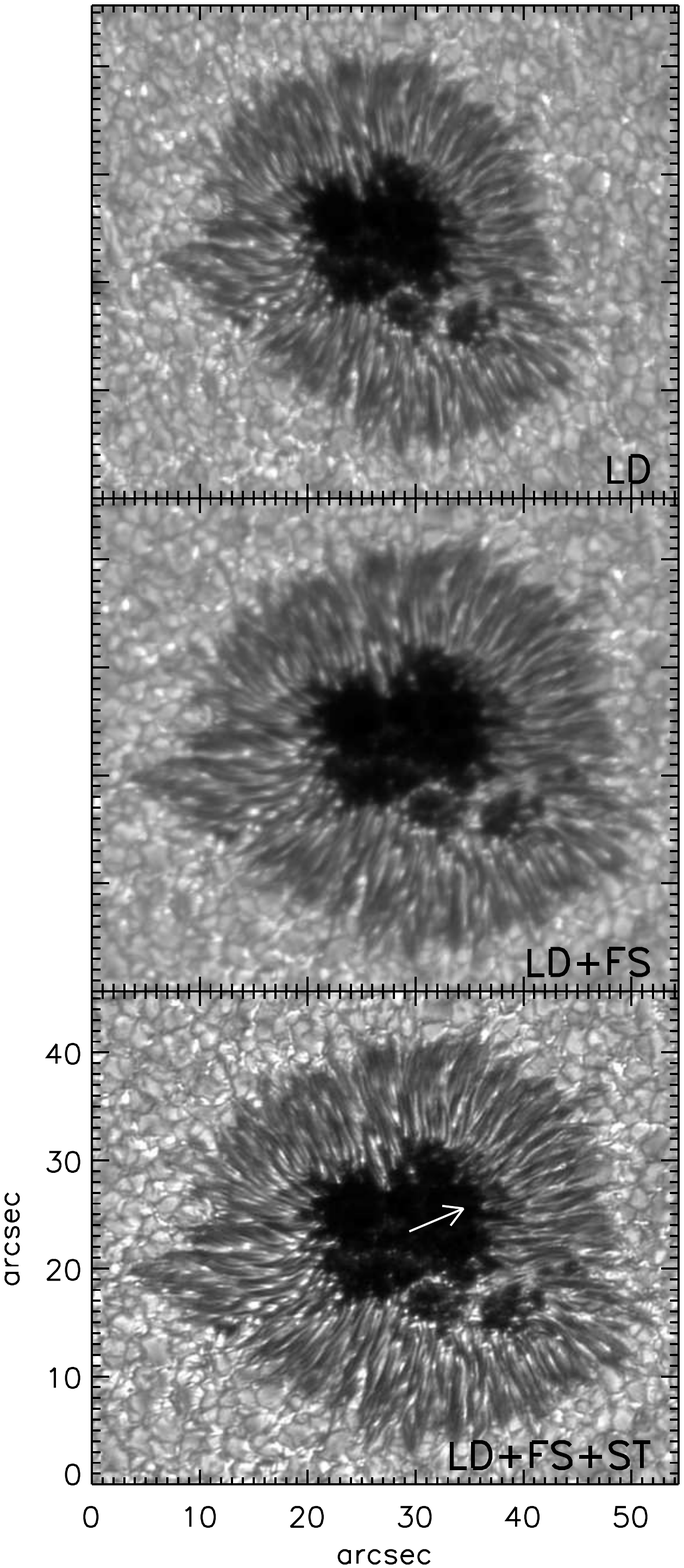}
}
\caption{G-band images of sunspot in NOAA AR 10961 located at S14  
E38 on June 28, 2007. The acronyms refer to the various
corrections applied to the Level-1 images: {\em{LD}} - limb darkening,
{\em{FS}} - foreshortening, {\em{ST}} - stray light. All images have
been scaled identically. Solar North points up and West is to the
right. The white arrow in the bottom panel points to disc center.}
\label{all3}
\end{figure}

\section{Observations and Data Processing}
\label{data}
\subsection{Imaging data}
We employ high resolution G-band filtergrams of NOAA AR 10961 recorded
from June 28 to July 5, 2007 by the Broadband Filter Imager (BFI) of
the Solar Optical Telescope \citep[SOT;][]{Tsuneta2008} on board {\em
Hinode}. The filtergrams have a spatial sampling of 0\farcs109 and
were taken in the 2$\times$2 binning mode with sizes varying from
1k$\times$1k to 2k$\times$1k.  The Level-0 G-band images were
corrected for dark current, flat field, and bad pixels using the {\em
fg\_prep} routine in SolarSoft. Additional corrections are described
below.

{\em Intensity Gradients:} The G-band data set covers the sunspot's
transit across the solar disc, from heliocentric angles ($\theta$) of
$\approx 50^\circ$ close to the east limb to $\approx 55^\circ$ near
the west limb. Consequently, intensity gradients in the
image arising from `limb darkening' \citep{Milne1921,Bohm1997} as well
as residual flat-field errors ought to be corrected.  A procedure for
correcting such intensity variations on {\em Hinode} G-band images has
been described by \citet{Tan2009}.  They selected a horizontal line
sampling the quiet Sun (QS) granulation and used a fifth degree
polynomial to fit the intensity as a function of $\mu =
\cos{\theta}$ and derive the limb darkening coefficients. A
different approach is described below. The filtergrams were first
normalized to the average QS intensity of the images taken close to
disc center ($\theta\sim12^\circ$). A 100$\times$100 pixel area was
then used to compute the mean QS intensity in all the images.  The
variation of intensity as a function of the $\mu$ corresponding to the
central pixel of the selected area is shown in the top panel of
Figure~\ref{limbpro}.  The coefficients given in the figure were
obtained from a second degree polynomial fit. Using these
coefficients, the intensity from each pixel was corrected as
$I_{\textrm{\tiny{obs}}}/I_{\textrm{\tiny{LD}}}$.  The effect of the
correction is illustrated in Figure~\ref{limbpro} for a horizontal cut
across the QS in a filtergram taken on June 28, 2007, when the sunspot
was located at a heliocentric angle of 38$^\circ$.

{\em Geometric Foreshortening:} Geometric foreshortening produces a
projected image of the sunspot as it traverses the solar disk from
East to West. A coordinate transformation described by
\citet{Gary1990} was performed to render the filtergrams into the
heliographic plane ($x_{\textrm{\tiny{H}}}$, $y_{\textrm{\tiny{H}}}$)
from the observed image plane ($x_{\textrm{\tiny{I}}}$,
$y_{\textrm{\tiny{I}}}$).  The middle panel of Figure~\ref{all3}
depicts the sunspot after correction for geometric foreshortening.

{\em Instrumental Stray Light: }The presence of stray light in the
{\em Hinode} broadband filtergrams arising from instrumental
scattering and its correction have been described by \citet{Shibu2009}
using transit observations of Mercury on November 8, 2006. In that
paper, the Level-1 images were deconvolved using a Point Spread
Function (PSF) consisting of a weighted linear combination of 4
Gaussians. As a result, the $rms$ contrast of bright points in the
quiet Sun improved by a factor of 1.7 in the G band (430 nm).  The
PSF derived by \citet{Shibu2009} is used here to remove stray light
from the G-band filtergrams assuming that the images corrected for
foreshortening are equivalent to those obtained with the telescope
pointing to disc center (the conditions of the transit observations).
The improvement in the image contrast after removal of stray light is
shown in the bottom panel of Figure~\ref{all3}, when the sunspot was
located at S14E38 on June 28, 2007. The implication of scattered light
on the fine structure of sunspots using {\em Hinode} filtergrams is
detailed in a separate work \citep{Rohan2012}.

\subsection{Spectropolarimetric data}
In addition to the G-band filtergrams, we also use observations taken
by the SOT spectropolarimeter
\citep[SP;][]{Lites2001,Ichimoto2008}. This instrument measured the
four Stokes profiles of the iron lines at 630~nm with a spectral
sampling of 21.55~m\AA\/, a pixel size of 0\farcs32, and an exposure
time of 1.6~s per slit position (fast map mode). The SP data were
corrected for dark current, flat field, thermal flexures, and
instrumental polarization using {\tt sp\_prep.pro}. The continuum maps
at 630 nm were also corrected for intensity gradients in a similar
manner as the G-band images. Maps of field strength, inclination and
azimuth were obtained from a Milne-Eddington inversion of the observed
Stokes profiles using the MERLIN code\footnote{The inversion results
can be found at the Community Spectropolarimetric Analysis Center,
\url{http://sot.lmsal.com/data/sot/level2hao\_new/}}. The procedure 
described by \citet{Crouch2009} was applied to the vector magnetograms
to resolve the 180$^{\circ}$ azimuth ambiguity. The resulting vector
magnetic fields were subsequently transformed to the local reference
frame.

\begin{figure*}[t]
\centerline{
\includegraphics[angle=180,width = 0.85\textwidth]{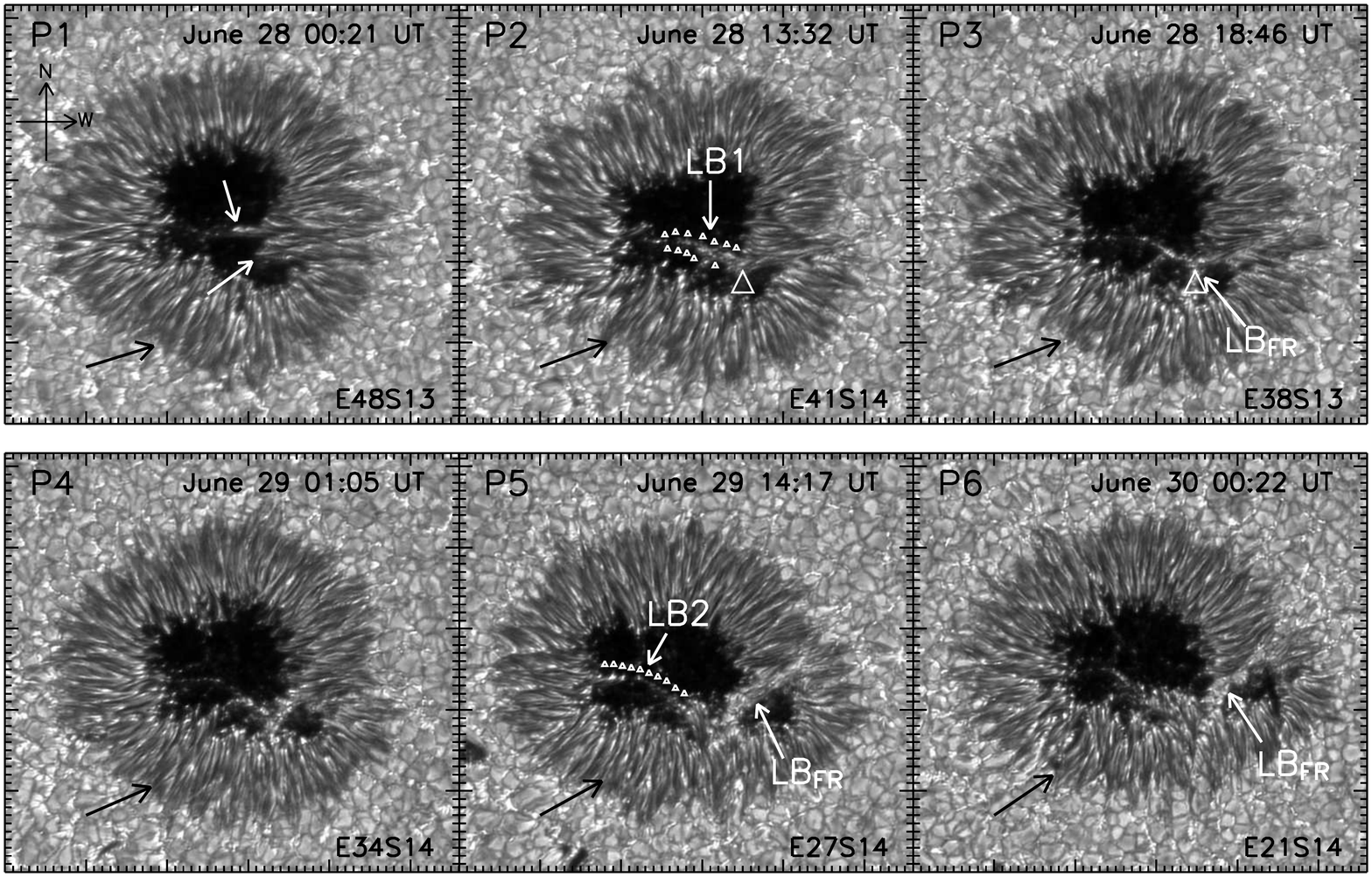}
}
\vspace{-26pt}
\centerline{
\includegraphics[angle=180,width = 0.85\textwidth]{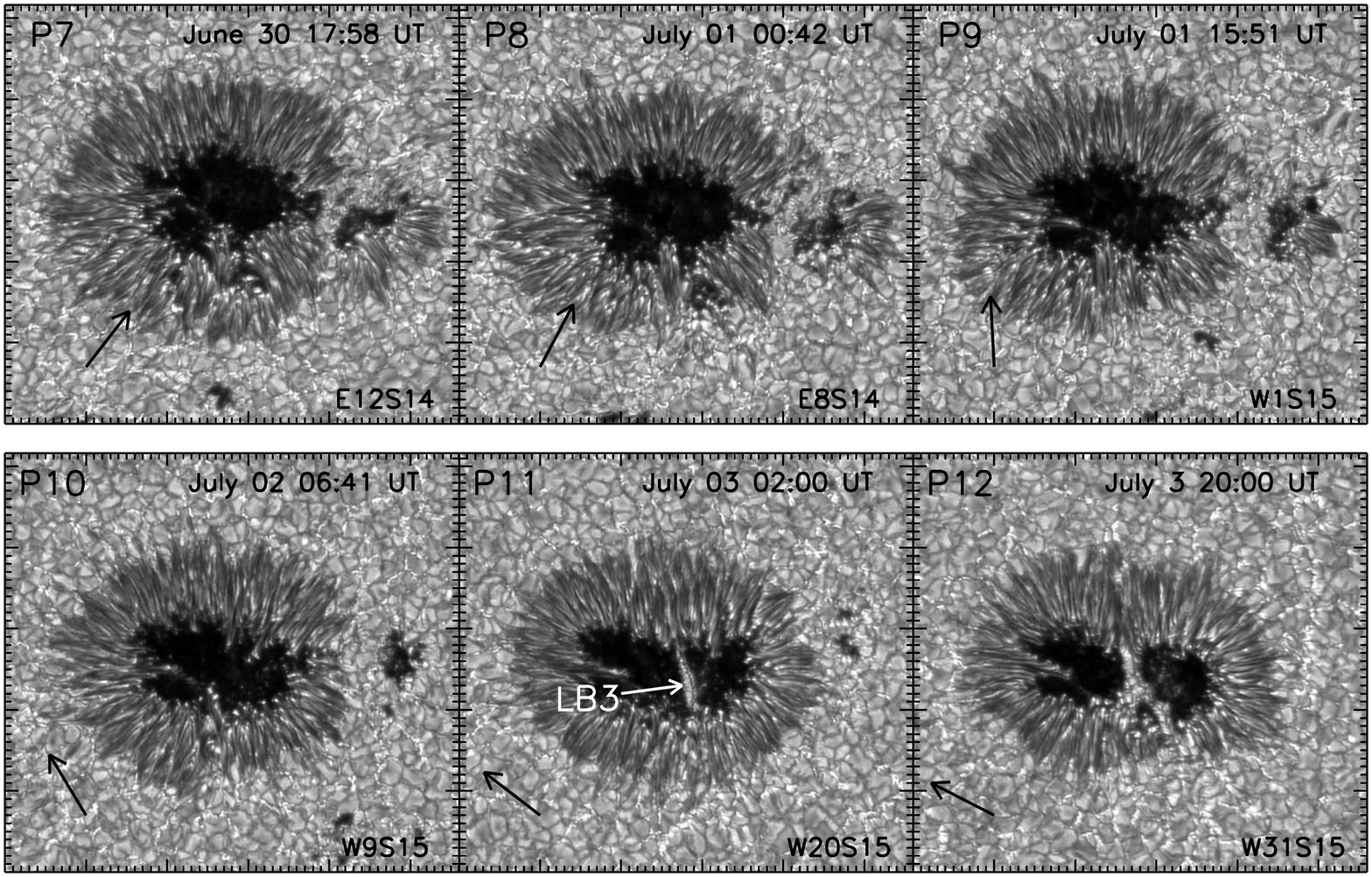}
}
\vspace{-20pt}
\caption{Evolution of sunspot in NOAA AR 10961 from June 28 to July 3, 2007. 
The position of the sunspot on the solar disc is shown at the bottom
of each panel. The image orientation is indicated in the legend at the
top left corner of panel P1. Each tickmark corresponds to
1\arcsec\/. The black arrow in the bottom left corner points to disc
center.See text for explanation of labels, symbols, and other arrows. }
\label{sunspot}
\end{figure*}

\section{Results}
\label{res}
\subsection{General Description of Sunspot Evolution}
\label{evol}

NOAA AR 10961 appeared close to the East limb during the early part of
June 25, 2007. The spot moved across the disc at about $14^{\circ}$S
and disappeared beyond the West limb on July 7.

Our data set begins on June 28 and shows a fairly regular sunspot
with a well developed penumbra (panel P1 of Figure~\ref{sunspot}).
The western side of the umbra-penumbra boundary shows two penumbral
filaments extending into the umbra (white arrows in P1). These
filaments progress deeper into the umbra eastwards giving rise to a
light bridge labeled `LB1' in P2 and outlined by tiny white triangles.
The apex of the large white triangle indicates a narrow umbral region
between the intruding filament and the southern umbra-penumbra
boundary (base of the triangle).  This location witnesses the
formation of another light bridge, labeled `LB$_{\textrm{\tiny{FR}}}$'
in the following panel P3. The time taken for the formation of
LB$_{\textrm{\tiny{FR}}}$ is approximately 5 hr. During this period,
LB1 becomes narrower and fainter in comparison to its newly formed
counterpart.  Panel P5 shows LB$_{\textrm{\tiny{FR}}}$ increasing
significantly in width from 485 km to 1295 km by the end of June 28,
nearly 20 hr after it was formed.  The panel also shows another light
bridge `LB2' (outlined with small white triangles), which can be
regarded as a successor of LB1. While LB1 was a penumbral structure,
LB2 predominantly consisted of umbral dots along its length. We choose
to treat LB1 and LB2 as distinct structures based on their different
morphology and time of formation.

During the early part of June 30 (P6 in Figure~\ref{sunspot}), the
symmetrical arrangement of the penumbral filaments near the southern
section of the umbra-penumbra boundary gets further disturbed. This
location coincides with one of the anchorage points of the light
bridge LB$_{\textrm{\tiny{FR}}}$. In addition, there are traces of
disruption in the penumbra with the absence of penumbral filaments in
the north-western sector of the sunspot, tracing outwards to the
photosphere from the entrance of LB$_{\textrm{\tiny{FR}}}$. The part
of the sunspot encompassed by the LB$_{\textrm{\tiny{FR}}}$ starts to
separate from the parent sunspot around 18:00 UT on June 30, as can be
seen in P7. At this stage, the two umbral boundaries marked by the
light bridge are separated by nearly 1.7 Mm and the morphology of the
light bridge resembles the quiet Sun photosphere. By the end of June
30, the fragmentation is complete with the sunspot and its fragment
being well separated. They remain within a distance of
10$^{\prime\prime}$ from each other.

The penumbra in the fragment gradually gets depleted as shown in P8 -
P9. By the early part of July 2, there are only minor traces of it in
the remnant pore (P10).  The penumbra in the parent sunspot gradually
becomes symmetrical and ordered with the exception of the region
closest to the fragment where the penumbral filaments are much shorter
in length and can be considered rudimentary. The decay of the fragment
and restoration of the penumbra in the parent are further elaborated
in Section~\ref{pen_reform}. The decaying pore is only seen as traces
of smaller fragments during the early part of July 3 (P11). While the
parent spot has a well developed penumbra uniformly structured all
around the umbra by July 3, a conspicuous light bridge `LB3' is seen
running from the northern to southern end of the umbra-penumbra
boundary, nearly splitting the umbra into two equal halves. LB3
progressively becomes broader and exhibits large granules along its
length. By June 5 the penumbra close to the southern part of LB3 gets
gradually depleted (P12).  The sunspot retains this configuration
until it advances over the Western limb on July 7, 2007.

\begin{figure}[t]
\centerline{
\includegraphics[angle=180,width = 0.55\textwidth]{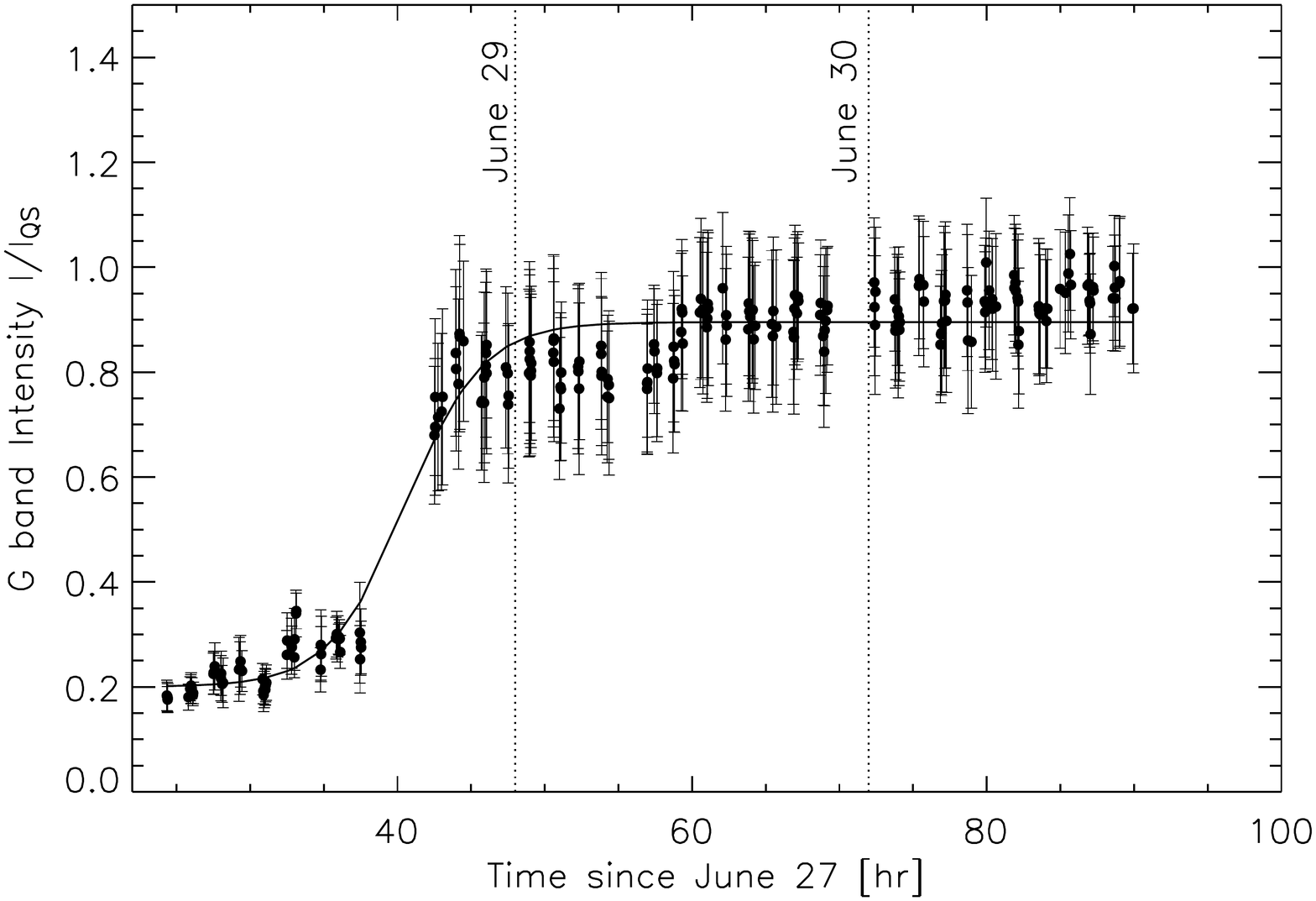}
}
\vspace{-10pt}
\caption{Evolution of G-band intensity in the light bridge. 
The {\em black circles} represent the mean G-band intensity of the
region where the light bridge was formed. The vertical bars indicate
the $rms$ fluctuations of the G-band intensity. The {\em solid} line 
is a Boltzmann sigmoid fit.}
\label{lb_int_curv}
\end{figure}

\subsection{Evolution of Light Bridge and Penumbrae}
\label{small}
This section describes the small-scale changes in the light bridge and
penumbrae associated with the evolution and fragmentation of the
sunspot.

\subsubsection{Formation and Evolution of LB$_{\textrm{\tiny{FR}}}$}
\label{lb_form}
The fragmentation of the sunspot is heralded by the formation of a
light bridge in the south-west sector of the sunspot umbra
(LB$_{\textrm{\tiny{FR}}}$, also called LB in this
section). The time sequence of G band images reveal that the LB
formed as a result of a slow incursion of penumbral filaments from
the north towards the southern umbra-penumbra boundary. This intrusion
is accompanied by an increase in the intensity by a factor of 2 over a
time duration of 12 hr starting at 1:49 UT on June 28. The speed of the filament motion 
into the umbra was estimated to be $\sim$0.08 km~s$^{-1}$ which is consistent with 
\citet{Katsukawa2007}. During the latter half of June 28, a coherent narrow 
LB is formed, isolating the smaller umbral core from the parent 
umbra (P3 of Figure~\ref{sunspot}). At this time, the LB consists of several bright, grain-like 
structures with a faint dark boundary separating individual cells (not shown). 
Intensities of these grains are comparable to those of bright penumbral 
grains at the southern end of the LB. We put an upper limit of $\sim$10 hr 
from the intrusion of the penumbral filament to the formation of 
the LB. The following 24 hr witnesses an increase in the width of the LB as
well as in the size of bright cells/grains on it, with the motion of the latter
closely resembling those of photospheric granules.

\begin{figure*}[!ht]
\hspace{-30pt}
\centerline{
\includegraphics[angle=0,width = 0.92\textwidth]{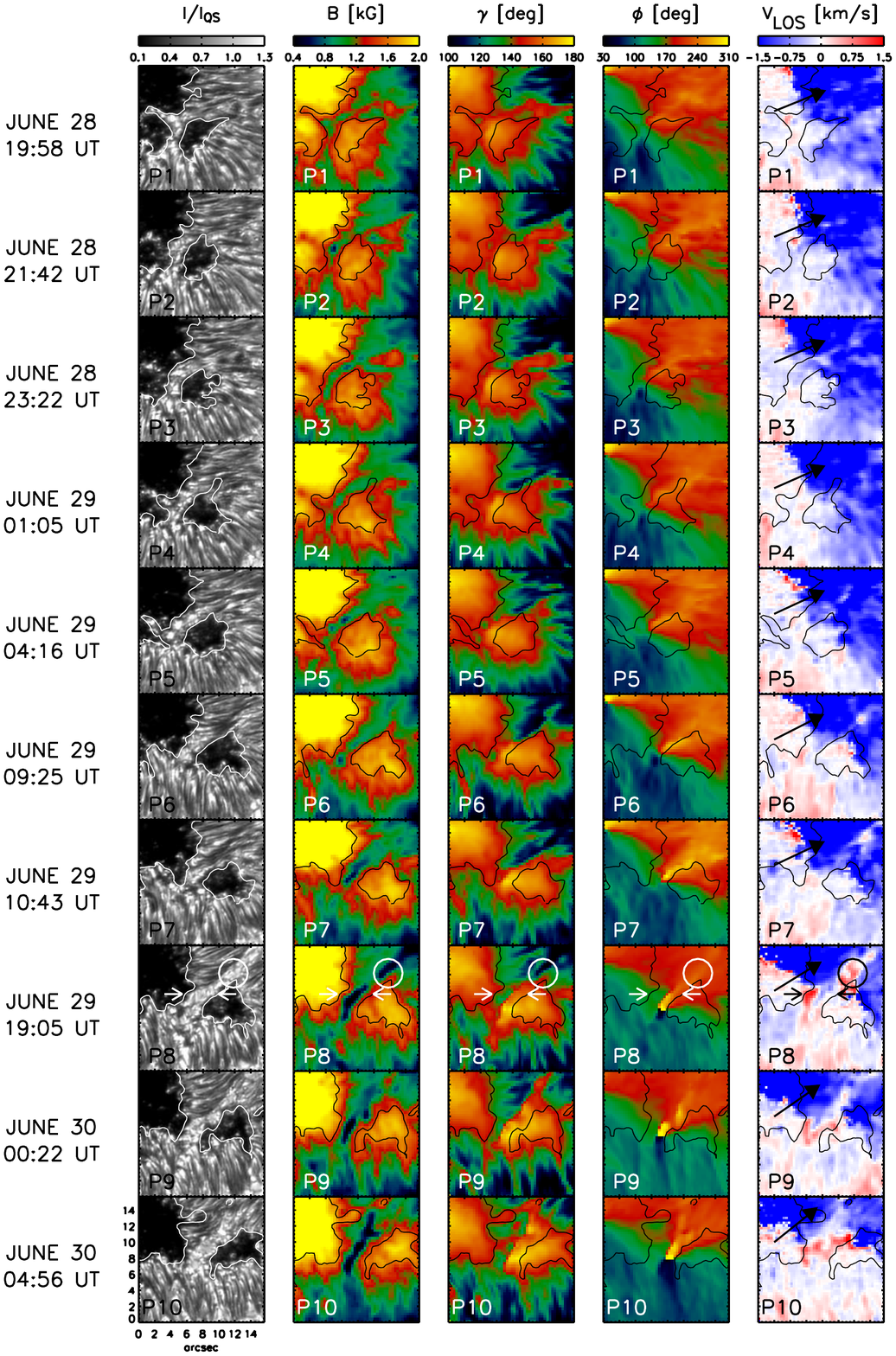}
}
\caption{Evolution of vector magnetic field and LOS velocity in the 
light bridge. {\em From left to right:} G-band intensity, field
strength, inclination, azimuth, and LOS velocity. All the maps have
been scaled as shown in their respective color bars. Solar East is to
the left and north is up. Inclinations and azimuths are expressed in
the local reference frame. Azimuths are measured counter-clockwise
from solar west (right). Thus, an azimuth of 0$^\circ$ implies a
magnetic field pointing along the positive x-axis. Blue/red correspond
to up/downflows in the last column. See text for explanation of arrow
head and region marked by circle in P8.}
\label{lb_maps}
\end{figure*}

The morphological transformation of the LB is accompanied by an
increase in its intensity. The temporal variation of the mean
intensity of the region where the LB formed is shown in
Figure~\ref{lb_int_curv}. The isolation of the smaller umbral core by
the LB is seen as a jump in the G-band intensity, increasing from 0.28
to 0.7 $I_{\textrm{\tiny{QS}}}$ in 4 hr. The steep increase is
followed by a more gradual rise touching near QS-photospheric
intensities close to midday of June 29. A Boltzmann sigmoid of the
form
\begin{eqnarray}
I(t) = a_0 + \frac{a_1 - a_0}{1 + \exp{((a_2 - t)/a_3)}} \nonumber
\end{eqnarray}
was used to fit the observed time variation of the mean G-band
intensity. The coefficients $a_0$ and $a_1$ correspond to the top and
bottom of the curve, respectively, $a_2$ is the time taken to reach
halfway between the top and bottom values, and $a_3$ is the slope of
the function (the larger its value, the shallower the curve). The best
fit yields $a_0 = (0.2 \pm 0.01) I_{\textrm{\tiny{QS}}}$, $a_1 =
(0.895 \pm 0.004) I_{\textrm{\tiny{QS}}}$, $a_2 = (16.58 \pm 0.14)$ hr, 
and $a_3 = (2.6 \pm 0.05) $ hr.

Figure~\ref{lb_maps} shows the temporal evolution of the G-band
intensity, magnetic field strength, field inclination, field azimuth,
and line-of-sight (LOS) velocity in the LB.  The contours represent
the umbral border and outline the LB.  The time sequence of
maps begins when the LB is already formed. While the main umbra
consists of fields in excess of 2 kG, the smaller umbral region
bounded by the LB is marginally weaker with an average field strength
of 1.5 kG. Compared to the adjacent umbrae, the field strength in the
LB is weaker by $\sim$500~G as seen in panels P1-P6. The magnetic
field in the LB is relatively inclined with respect to the vertical,
reaching values of 120-150$^\circ$.  
As a consequence of the smaller
core being bounded by the LB, one azimuth center is located close to
the central part of the LB on its western edge. 
The field is predominantly oriented across the LB. This is different 
to the configuration found in other LBs (e.g., \citet{Rohan2008}). The LOS
velocity in the LB (panels P1--P6) shows weak upflows in the range 0.1--0.35
km~s$^{-1}$, as well as downflows of 0.25--0.57 km~s$^{-1}$. These
flows are quite weak in comparison to the penumbra whose disc and limb
sides show velocities greater than $\pm$1km~s$^{-1}$.

\begin{figure}[!h]
\centerline{
\hspace{15pt}
\includegraphics[angle=180,width = 0.5\textwidth]{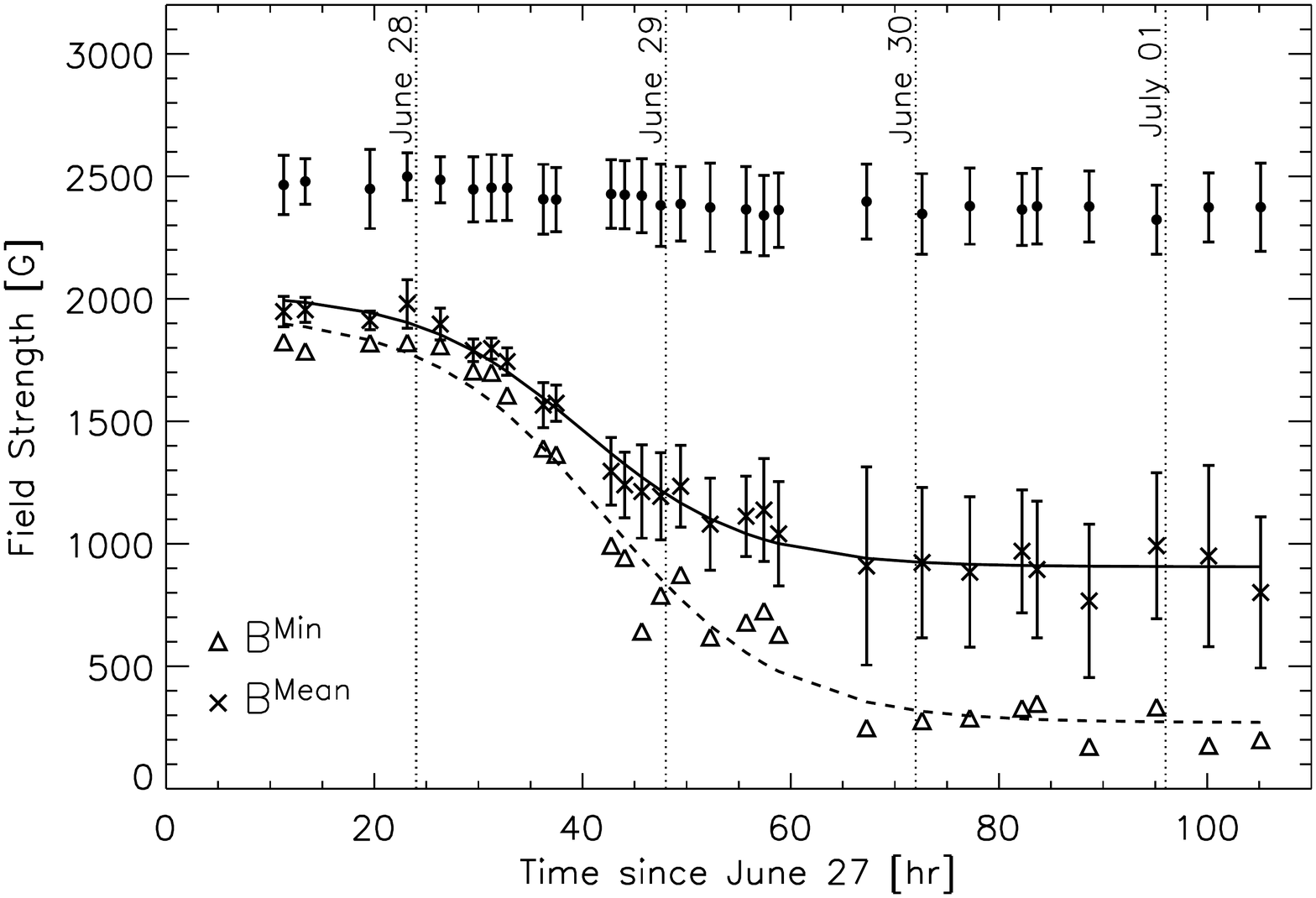}
}
\vspace{-10pt}
\caption{Evolution of field strength in the light bridge. 
The {\em crosses} and {\em triangles} represent the mean and minimum
field strength of the region where the light bridge was formed. The
vertical bars represent the $rms$ fluctuations. The {\em solid} and
{\em dashed} lines are Boltzmann sigmoid fits to the observed mean and
minimum field strengths. The {\em black circles} correspond to the
mean umbral field strength close to the LB.}
\label{field_evol}
\end{figure}

Nearly 20 hr after its formation, the field strength along the axis of the LB
drops to 650 G (P7). Panels P8--P10 indicate the following: i) the penumbra
close to the northern entrance of the LB appears to be sidelined to
the eastern edge while bright photospheric material is seen intruding
from the periphery of the spot all the way to the LB, ii) the minimum
field strength in the LB reduces to 300 G, iii) a narrow upflowing
lane is seen along the central axis of the LB with downflows of 1.5
km~s$^{-1}$ adjacent to it, and iv) strong downflows are also seen in
photospheric regions of the LB near penumbral filaments (circle in
P8).

\begin{figure*}[!ht]
\centerline{
\hspace{-10pt}
\includegraphics[angle=0,width = 0.53\textwidth]{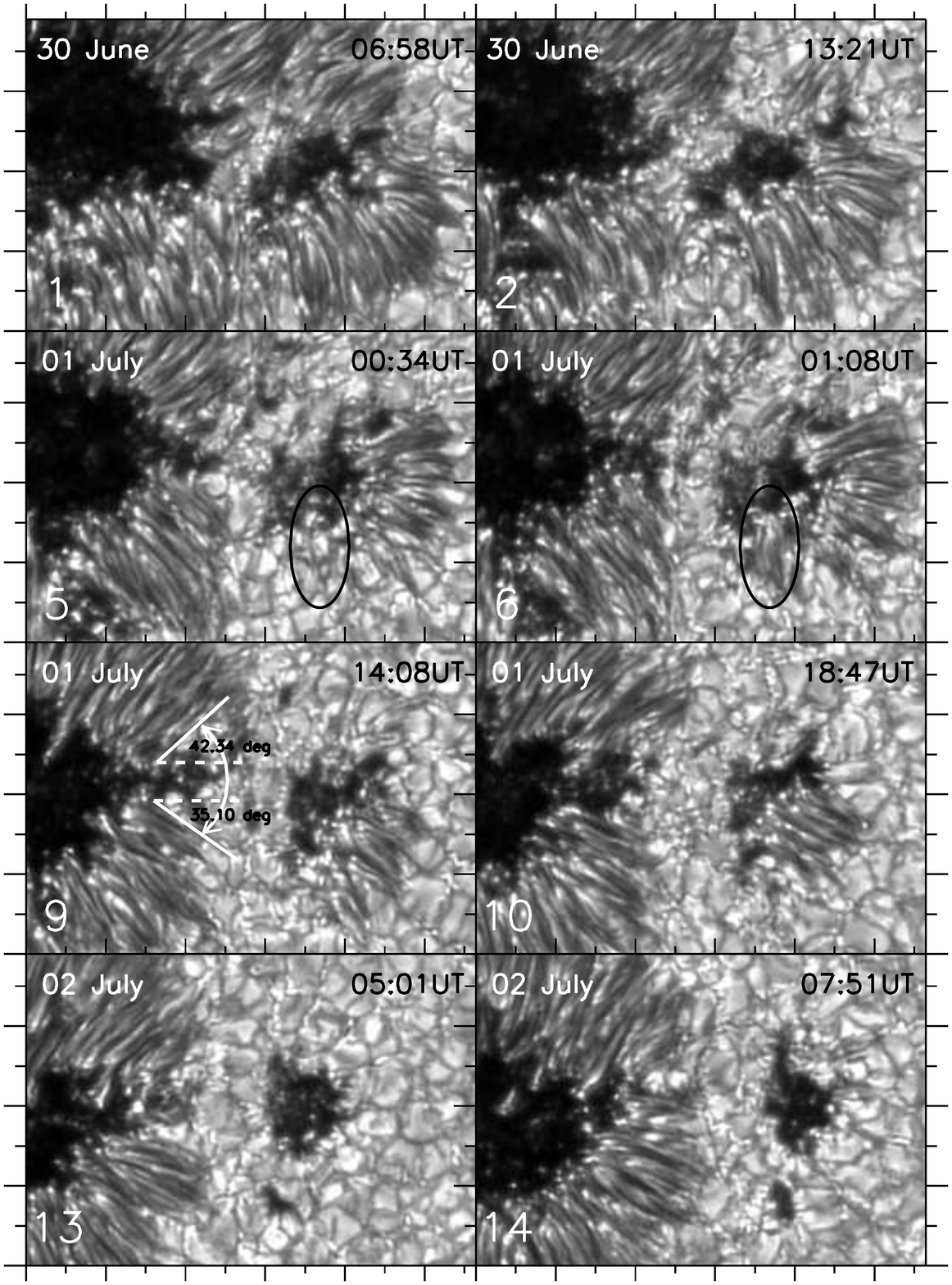}
\hspace{-32pt}
\includegraphics[angle=0,width = 0.53\textwidth]{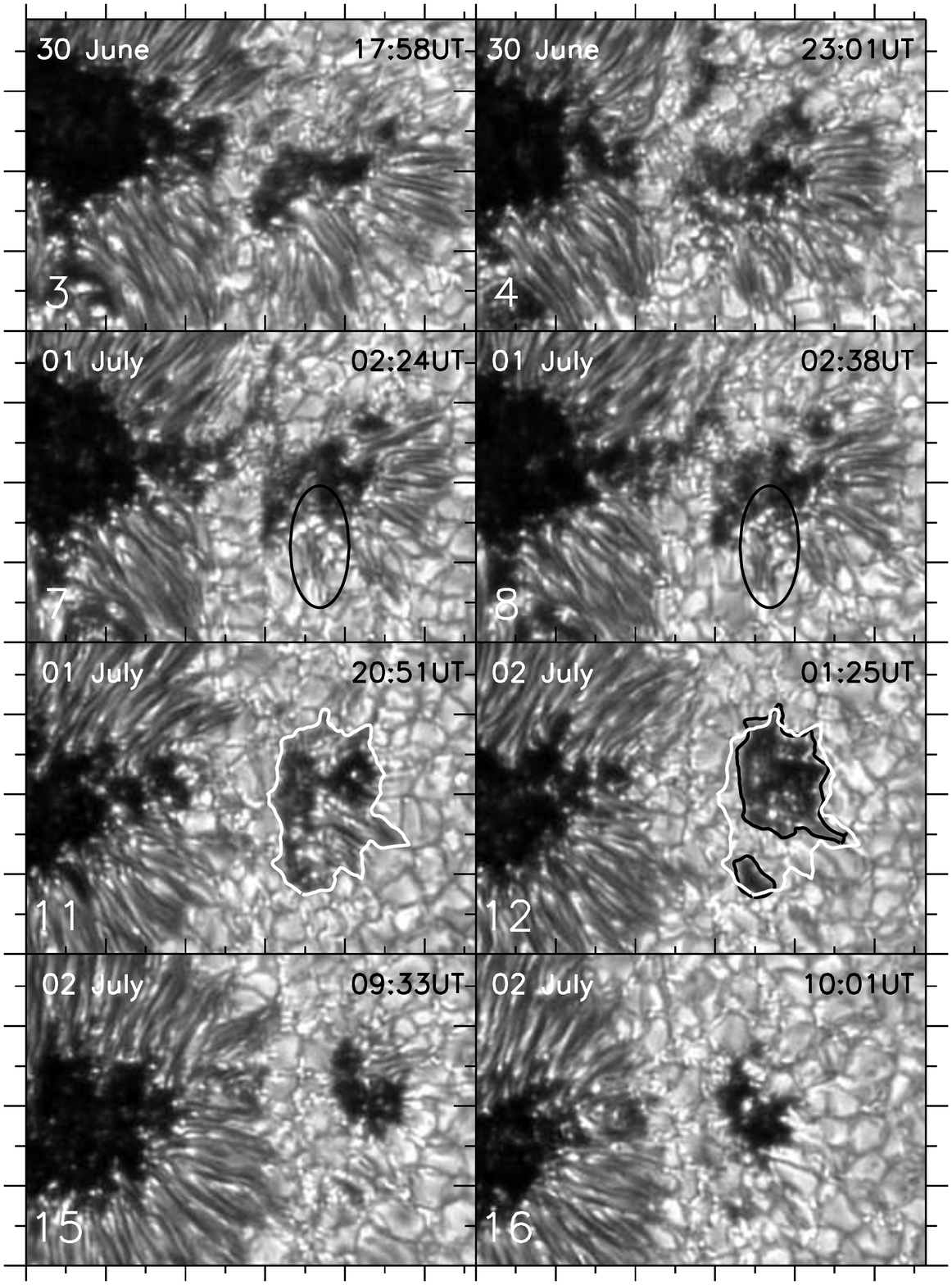}
}
\vspace{-46.5pt}
\centerline{
\hspace{-10pt}
\includegraphics[angle=0,width = 0.53\textwidth]{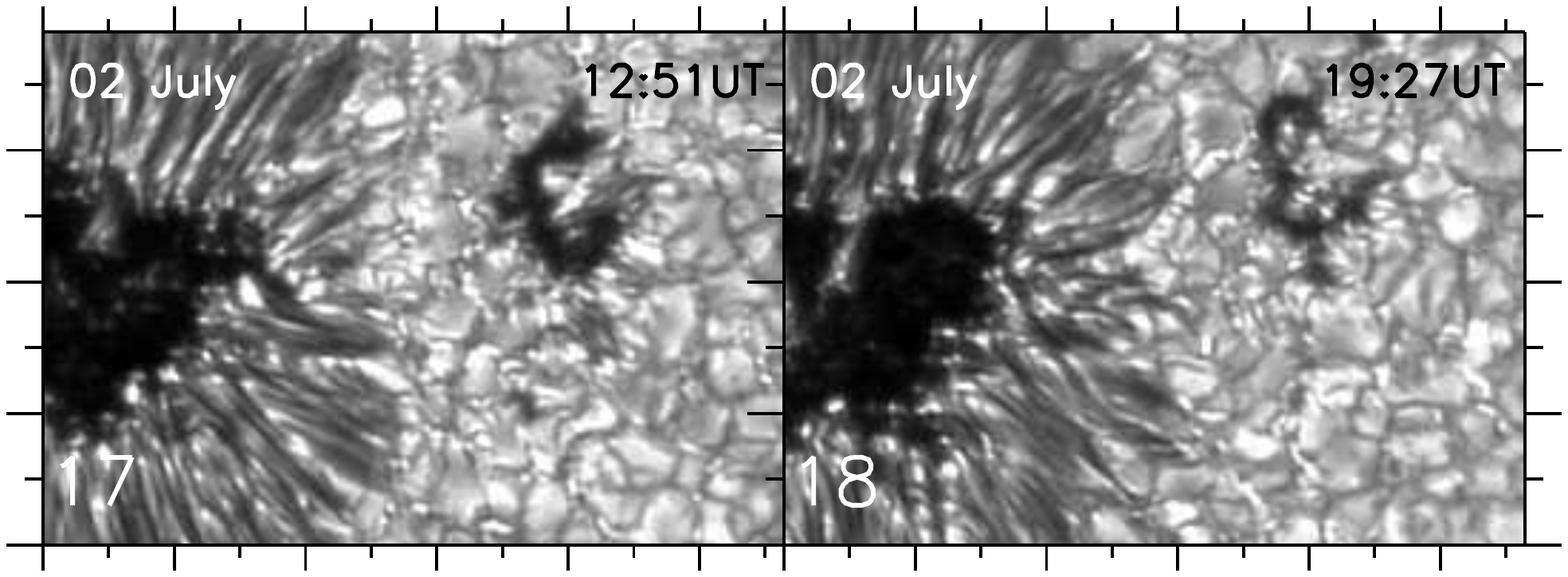}
\hspace{-32pt}
\includegraphics[angle=0,width = 0.53\textwidth]{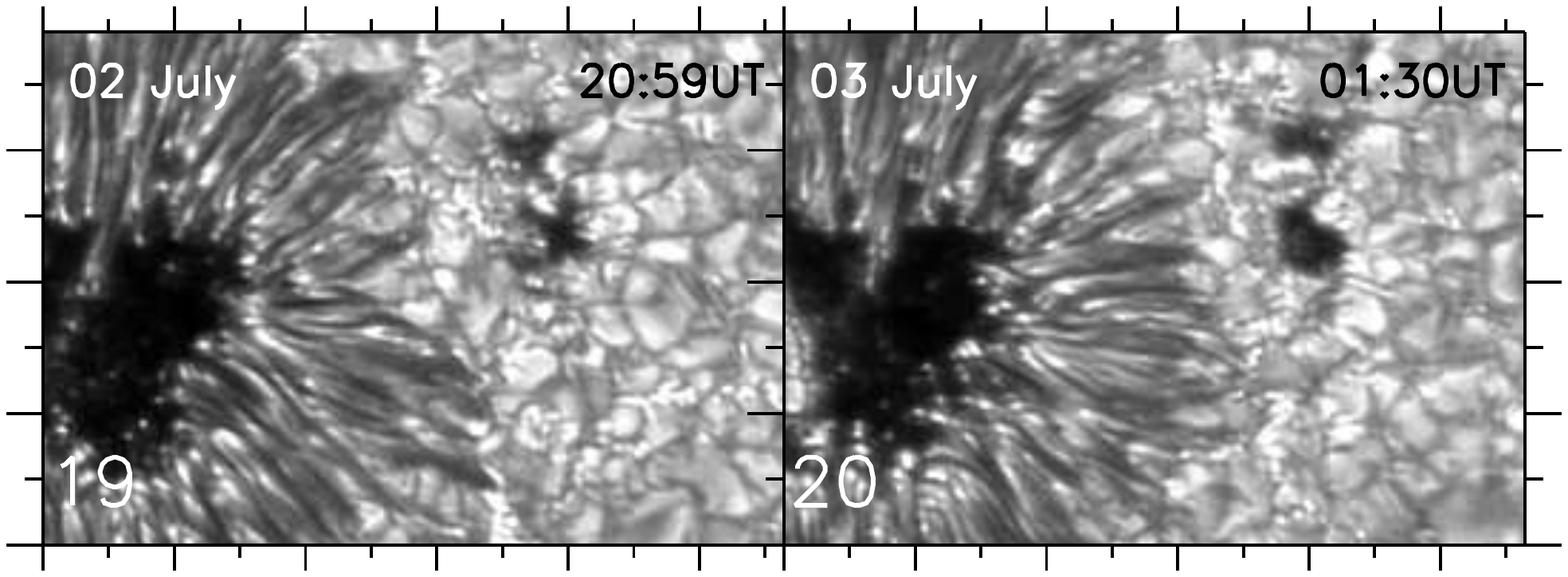}
}
\vspace{-290pt}
\caption{Decay of fragment and formation of penumbra in the parent sunspot 
in the post fragmentation phase. The ellipse in panels 5 to 8 depicts transient 
penumbra in the decaying fragment. The white contour in panels 11 and 12 
outline the fragment on July 1, 20:51 UT, while the black contours in panel 
12 mark the pore and the magnetic patch on July 2, 01:25 UT. Each major tickmark corresponds 
to 5\arcsec\/.}
\label{combi}
\end{figure*}

The LOS velocity and the field strength along a cut (arrow head across
the LB in P8) indicate a reduction in field strength that coincides
with the upflowing lane and is flanked by downflows on either
side. The amplitude of Stokes $V$ in the pixels exhibiting the upflows
is weaker than in the downflows. Downflows of up to 5~km~s$^{-1}$ are
seen in the region marked with a circle in P8. The Stokes $V$ profiles
emerging from those pixels show an extended hump characteristic of
very large velocities. The magnitude of the downflows was derived from
a two-component SIR inversion \citep{Ruiz1992} with height-independent
parameters (except for temperature). The fill fraction of the fast
component was estimated to be 16\%.

The evolution of the mean and minimum field strengths in the region
where the LB formed is depicted in Figure~\ref{field_evol}. The
weakening of these quantities is similar to the trend seen in the
G-band intensity. A Boltzmann sigmoid fit to the field strength
reveals that the lower knee of the curve coincides with the weak
fields detected in the LB on June 28, 19:05 UT (P8 of
Figure~\ref{lb_maps}).  The morphology of the LB from this point
onwards also appears distinct from the nearby penumbral features.  The
time taken for the field strength (and the continuum intensity) to
reach halfway between the bottom and top values is $\sim$16 hr,
starting from June 28. By comparison, the field strength of the 
neighbouring umbral region remained unaffected during the evolution 
and morphological transformation of the LB. With a significant field 
weakening in the LB and the onset of convection that is revealed by 
the upflows and downflows, the sunspot split close to midday of 
June 30, nearly 45 hr after the formation of the LB. The following 
section describes the evolution of the fragment and its 
subsequent decay.

\subsubsection{Evolution of Fragment}
\label{fra_evo}
Figure~\ref{combi} depicts the temporal evolution of the sunspot
fragment and the disappearance of its penumbra. The first panel shows
the fragment during the early part of June 30 when it was still part
of the parent sunspot.  The angular span of the penumbra around the
smaller umbral core which broke away is $\approx$120$^\circ$.
Fragmentation commences with the depletion of penumbra close to the
anchorage points of the light bridge. The penumbra is seen to reduce
symmetrically on either side of the fragment, disappearing clockwise
and counter-clockwise in the northern and southern penumbra,
respectively.  The angular span reduces to half its initial value
within nearly 2 hours between June 30, 23:01 UT and July 1, 00:34 UT.
The sector-wise disappearance of the penumbra is almost a reversal of
the formation process observed by \citet{Schlichenmaier2010}.  There
is no major trace of the penumbra by July 2, with an isolated pore
being the remnant of the fragment, although individual strands of the
penumbra can be spotted around the pore (panel 12).  There is also a
reduction in the mean width of the fragment penumbra, which decreased
from $6\farcs0 \pm 0\farcs5$ on June 30 to $4\farcs6 \pm 0\farcs5$ on
July 1. The above values correspond to a 24 hr average.

\begin{figure}[!h]
\centerline{
\includegraphics[angle=180,width = 0.45\textwidth]{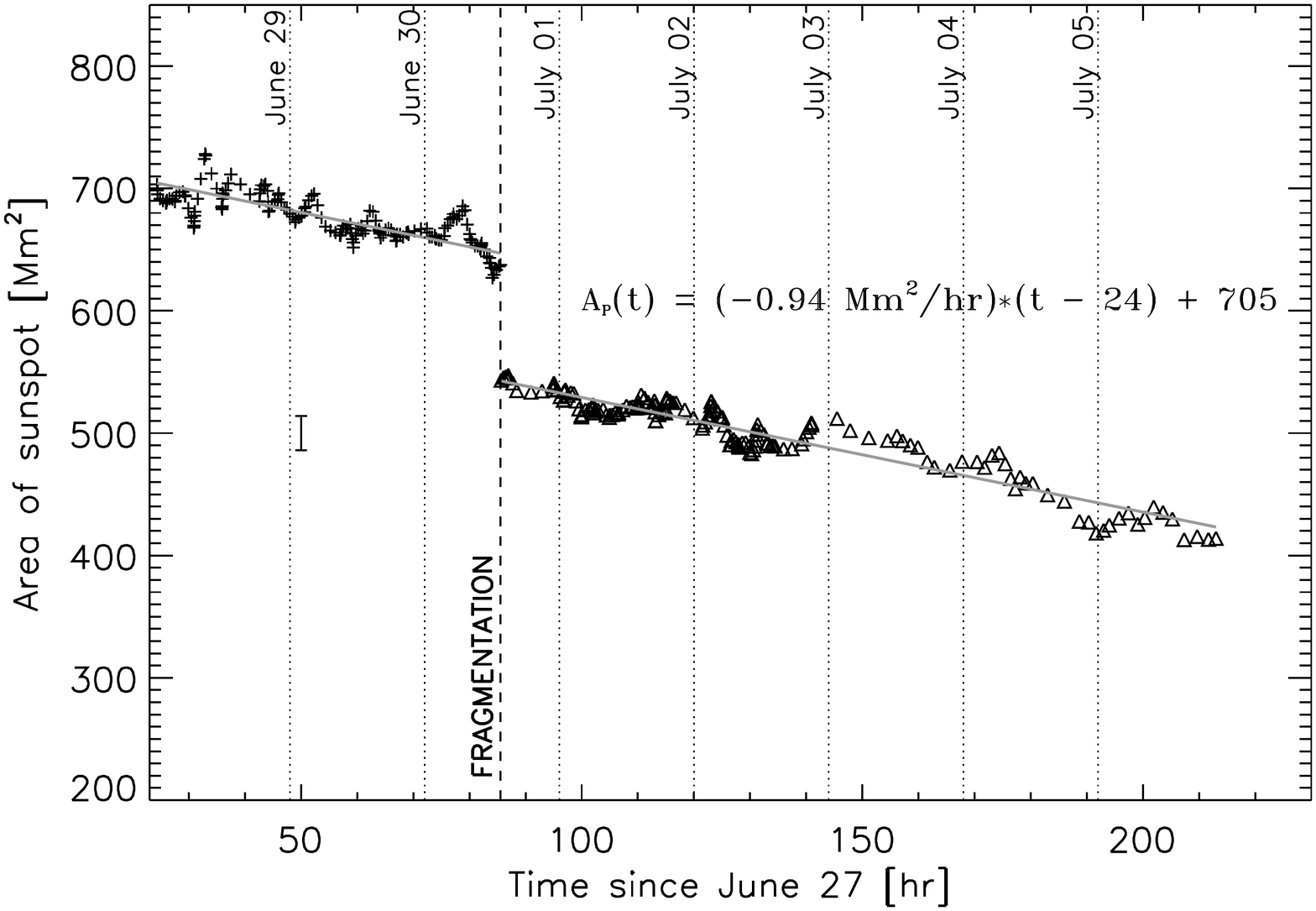}
}
\vspace{-10pt}
\centerline{
\includegraphics[angle=180,width = 0.45\textwidth]{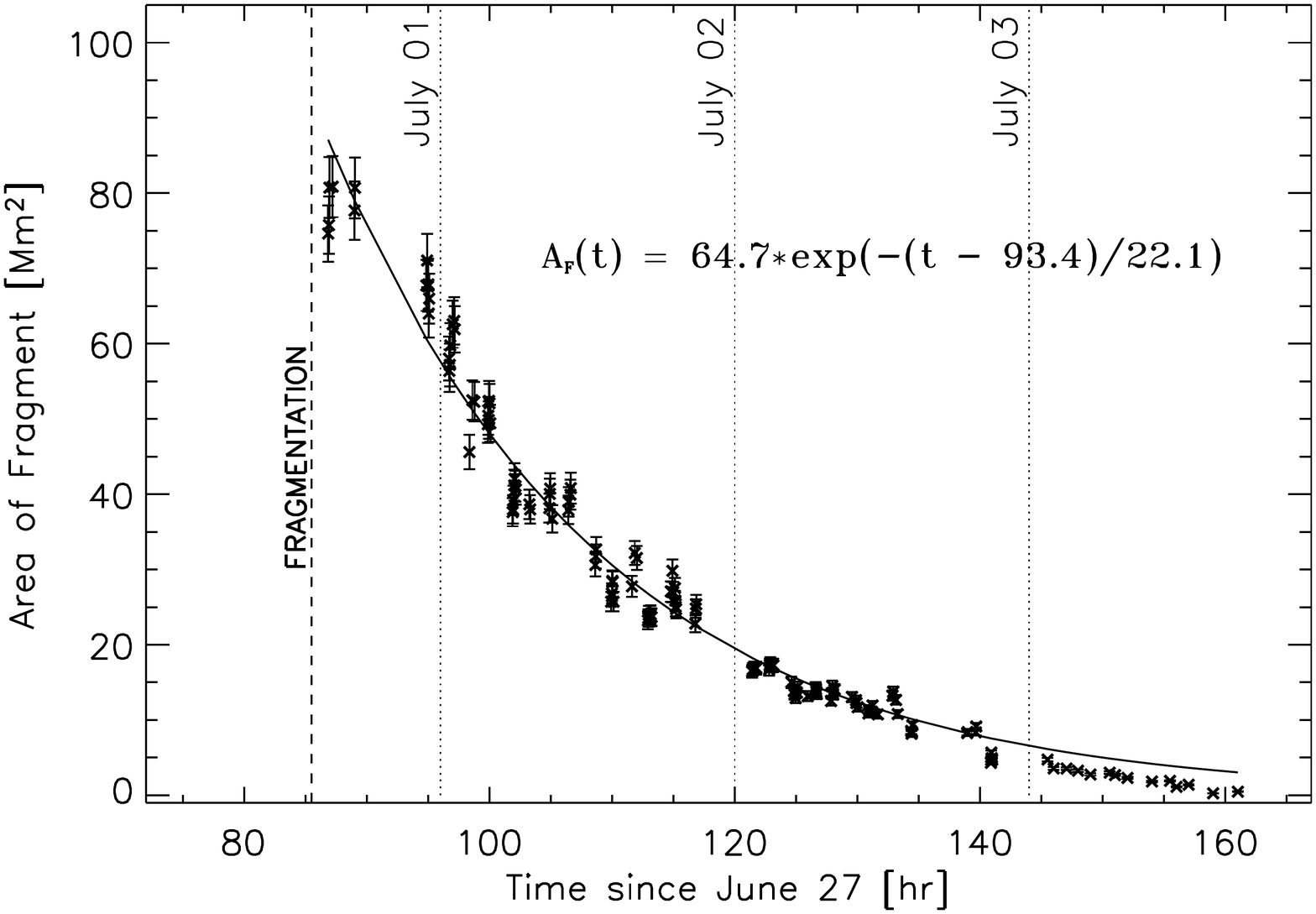}
}
\caption{Decay of NOAA AR 10961. {\em Top:} Sunspot area before
fragmentation ({\em plus symbols}) and after fragmentation
({\em triangles}). The area was computed from the G-band images. Rapid 
changes in area are neglected by smoothing the data by a 3 point 
average. The typical error in the area estimation is indicated at 
data point (50, 500). The {\em solid grey} line is a linear fit to the 
data points. {\em Bottom:} Decay of sunspot fragment. The {\em
cross} symbols show the fragment area as a function of time. The {\em
solid} line is an exponential fit to the curve. The vertical bars
indicate the error arising from a $\pm$5\% uncertainty in the
intensity contour level.}
\label{deca}
\end{figure}

At the time of fragmentation on June 30, 13:39 UT, the fragment has an
area of 92 Mm$^2$ and decays exponentially with a time constant of
$22.1 \pm 0.2$ hr (bottom panel of Figure~\ref{deca}).  Panels 11 and
12 of Figure~\ref{combi} depict the transition of the fragment to
a pore devoid of a penumbra. This occurs between July 1, 20:51 UT and
July 2, 01:25 UT. The white contour in panel 11 corresponds to the
penumbra-QS boundary. It has also been overlaid in panel 12, which
shows a large pore accompanied by a tiny magnetic patch located near
the bottom edge of the white contour. The area of the fragment on July
1, 20:51 UT is 33 Mm$^2$, with umbral and penumbral contributions of
14 and 19 Mm$^2$, respectively.  The pore and the patch lie within the
white contour representing the fragment 4.5 hr earlier. The total area
of the pore and its satellite fragment is $\sim$21 Mm$^2$ at that
moment. If the decrease in the area is attributed to the disappearance
of the penumbra and if the natural decay rate of the fragment is
neglected, then the pore and the patch ought to have decayed to at
least 14 Mm$^2$, which is inconsistent with the measured value. This
suggests that the pore gets replenished with a fraction of the earlier
decaying penumbra. During penumbral formation, the area of the umbra
remains constant and the growth of the penumbra alone accounts for the
increase in spot area \citep{Schlichenmaier2010,Rezaei2012}.  The process we have
witnessed suggests a different behavior for penumbral decay.

\begin{figure}[!h]
\centerline{
\includegraphics[angle=180,width = 0.5\textwidth]{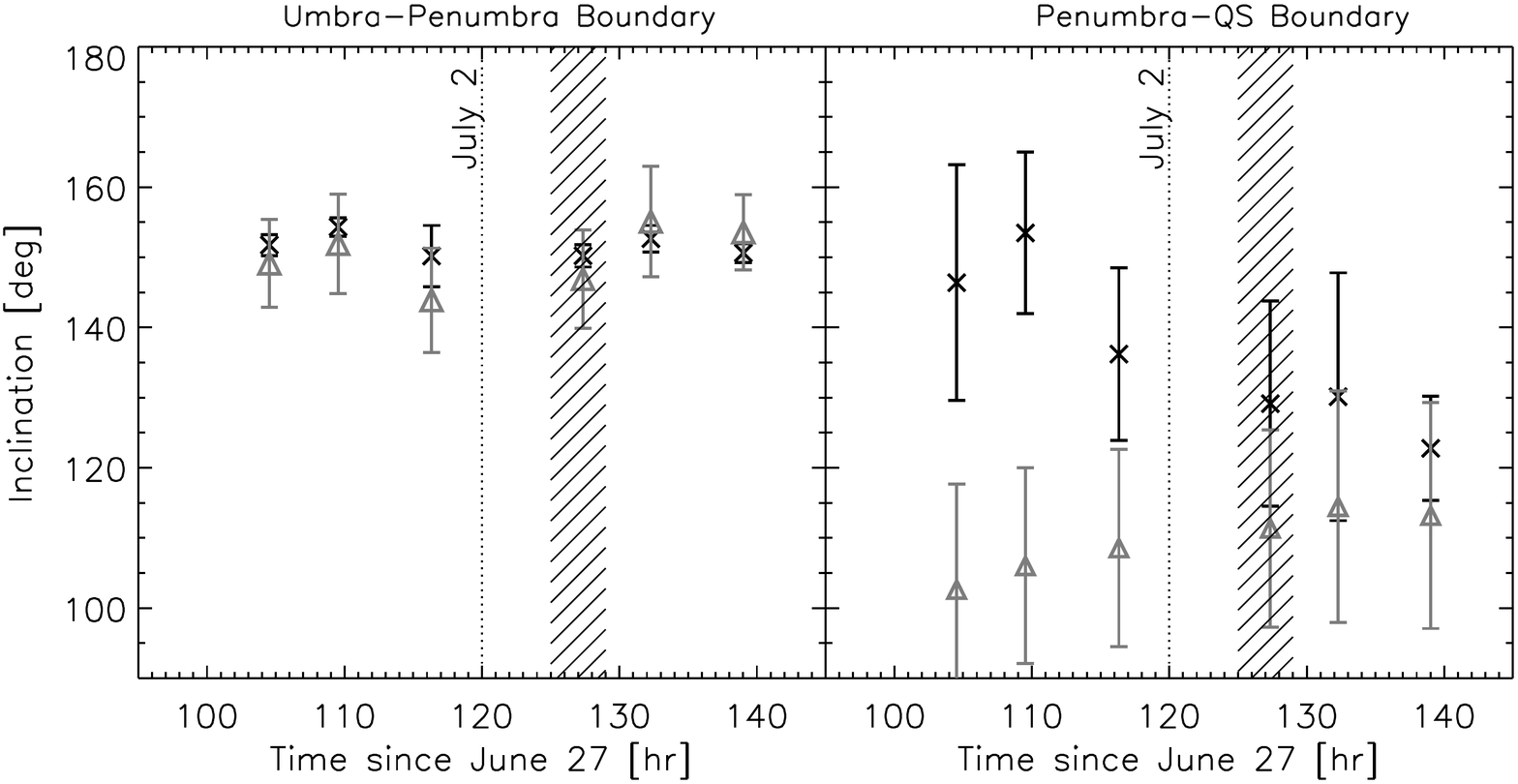}
}
\vspace{-45pt}
\caption{Variation of field inclination at the inner and outer penumbral 
boundaries.  {\em Left:} Time variation of the mean inclination at the
umbra-penumbra boundary. The {\em crosses} and {\em triangles}
correspond to the fragmentation site and the rest of the spot,
respectively. The vertical error bars represent $rms$ values. The
hatched region denotes the interval during which formation of penumbra
was seen in the sunspot at the location of fragmentation. {\em Right:}
Same, but for the outer penumbral boundary.}
\label{inc}
\end{figure}

One also finds evidence of transient penumbrae that appear and
disappear close to the southern edge of the fragment. This is
illustrated in panels 5 to 8 of Figure~\ref{combi} with ellipses, 
which shows the appearance of rudimentary penumbrae in locations where 
photospheric granulation was observed earlier (panels 5 and 6).  Such 
structures do not form stable penumbral filaments.  Transient penumbrae 
can develop quite rapidly over periods of $\sim$30 min, but their 
disappearance is more gradual and occurs over periods of 1.5 hr or more. One 
can identify two distinct penumbral filaments at the position marked by 
the ellipse in panel 5 of Figure~\ref{combi}. Panel 7 shows several bright 
granules clustered together at the edge of the umbra-QS boundary with intensities
comparable to those of regular penumbral grains. In a span of 15 min,
a single penumbral filament is observed in this region. The
rudimentary penumbra is not visible after July 1, 5:43 UT.  Similar
structures have been reported by \citet{Schlichenmaier2010} during penumbral 
development, so they are not exclusive to penumbral decay.

\begin{figure}[!h]
\centerline{
\includegraphics[angle=180,width = 0.45\textwidth]{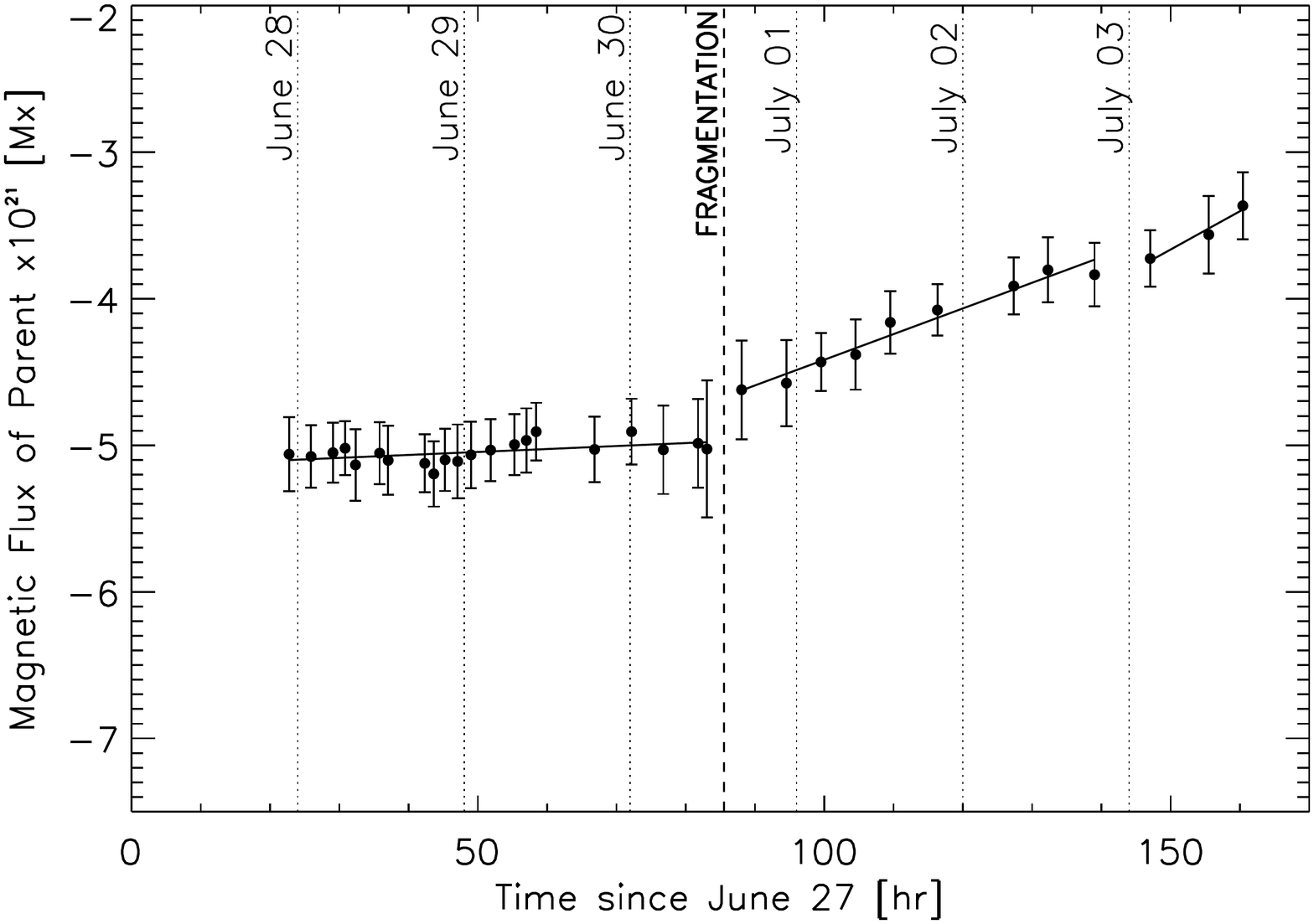}
}
\vspace{-10pt}
\centerline{
\includegraphics[angle=180,width = 0.45\textwidth]{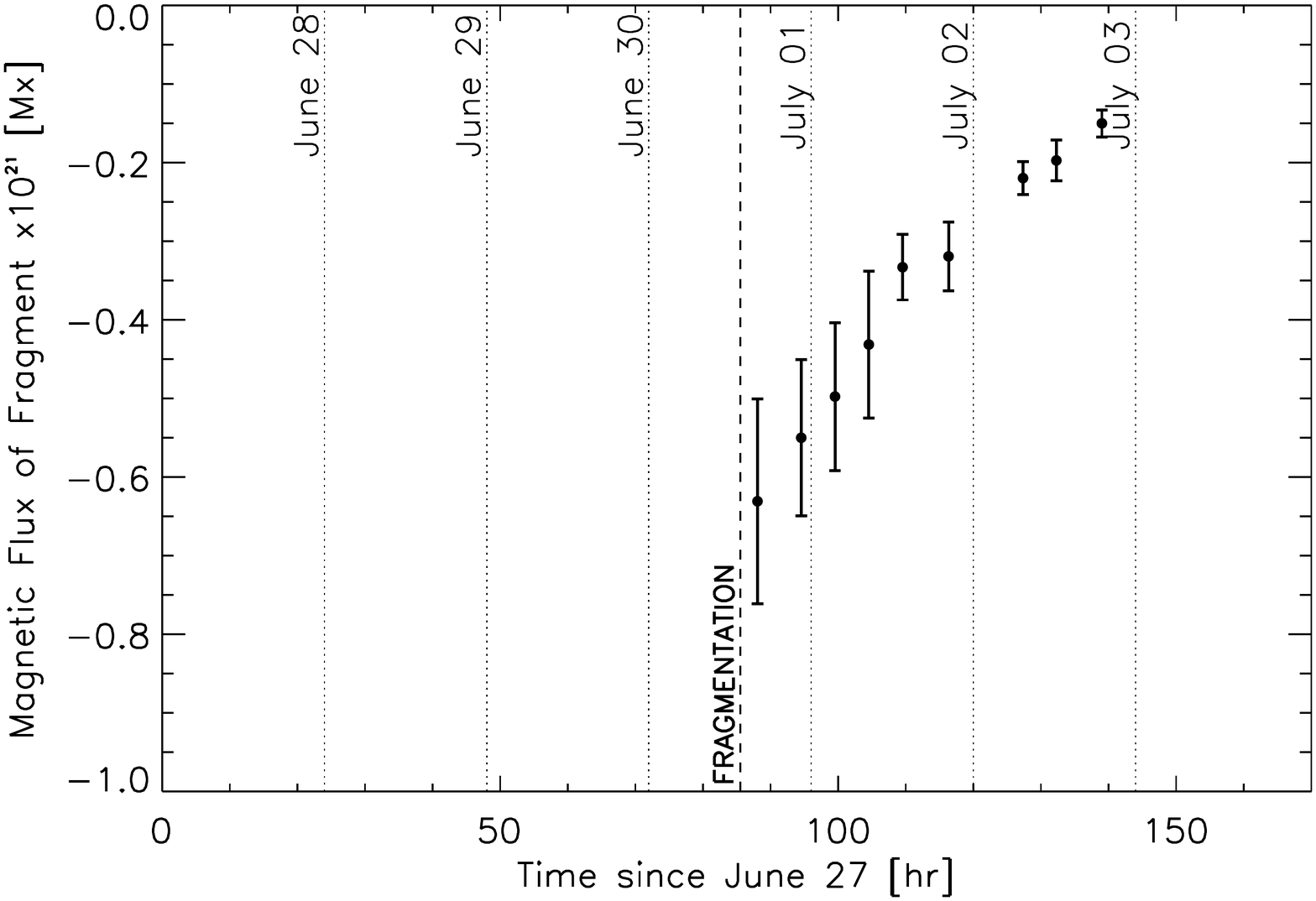}
}
\caption{Evolution of the magnetic flux of NOAA AR 10961 derived from 
SP magnetograms. {\em Top:} Magnetic flux of the 
parent. The error bars correspond to an uncertainty of 5\% in the 
iso-intensity contour separating the parent/fragment from the QS.
{\em Bottom:} Same as above, but for magnetic flux of the fragment.}
\label{deca_flux}
\end{figure}

\subsubsection{Restoration of Penumbra in Parent Sunspot}
\label{pen_reform}
The separation of the fragment renders a discontinuity in the
azimuthal arrangement of the penumbra in the parent sunspot which coincides with
LB$_{\textrm{\tiny{FR}}}$. Panels 9 to 20 of Figure~\ref{combi} show 
the intervening photosphere between the fragment and the parent sunspot. The angular
span devoid of the penumbra in the parent is $\sim$80$^\circ$ (panel 9
of Figure~\ref{combi}). The appearance of a complete penumbra
in this region is observed only towards the end of July 2, nearly 40
hr after the fragmentation.  Furthermore, the length of the penumbral
filaments in this region is about 1.5 times shorter than the average
filament elsewhere in the sunspot. There is also a strong anti-correlation
between the area of the fragment and the width of the parent's penumbra
closest to the fragment. This would suggest that the presence of the 
fragment hinders the formation of the penumbra in the parent spot.

To determine the threshold of the magnetic field inclination 
and how it relates to the regeneration of the penumbra in the parent, we 
compare the field inclination ($\gamma$) at the location
of fragmentation with the rest of the sunspot having a regular
penumbra, both for the umbra-penumbra boundary (UPB) and for the
penumbra-QS boundary (PQB). 
The inclination in the region of fragmentation is calculated 
along the portion of the UPB and PQB
intensity contours which is closer to the fragment. The left panel of
Figure~\ref{inc} shows the time evolution of $\gamma$ at the UPB. The
mean inclination is $\sim$150$^\circ$ at the location of fragmentation
as well as the rest of the sunspot. However, the right panel shows the
field becoming more inclined at the PQB, from 146$^\circ$ on July 1,
8:30 UT to 130$^\circ$ on July 2, 7:21 UT. The average value of
$\gamma$ in the rest of the sunspot shows typical values of
100--115$^\circ$, i.e., 10--25$^\circ$ from the horizontal (the spot
is of negative polarity), which is consistent with \citet{Jurcak2011}. 
The increase in inclination coincided with
the growth of the penumbra that occurred between 5:00-9:00 UT on July 2
(refer panels 13 to 16 of Figure~\ref{combi}). Thus the radial
width of the penumbra increased when the average field inclination at
the border of the sunspot dropped below 130$^\circ$ (i.e., when the
field became within 40$^\circ$ of being purely horizontal).

\subsection{Area Decay of Sunspot and Fragment}
\label{area}
Since the sunspot loses a significant area during 
the fragmentation process, it is necessary to compare its area 
and flux decay rates with that of the fragment.
Estimation of the sunspot area relies on determining the intensity level
separating the penumbra/umbra from the QS. Once the intensity at the
interface is known, the area can be calculated from the total number
of pixels within the corresponding contour. The intensity of the
penumbra-QS boundary was determined with the help of the cumulative
histogram method of \citet{Pettauer1997}, which yielded a value of
0.925 for the G-band time sequence.
In order to obtain a smooth contour, the filtergrams were filtered
using a 7$\times$7 pixel boxcar.

The top panel of Figure~\ref{deca} shows the decay of the sunspot's
area with time. The sunspot area decreased from 697
Mm$^2$ at 00:21 UT on June 28 to 636 Mm$^2$ at 13:39 UT on June
30. A linear fit to the data points yields a decay rate of 
$-22\pm1.2$~Mm$^2$/day, or $-7\pm 0.4$~MSH/day\footnote{1 MSH = 6.3 arcsec$^2$}
with a jump of 93 Mm$^2$ at the time of fragmentation.

The bottom panel of Figure~\ref{deca} shows the area decay of the
fragment. In contrast to the parent, the decay appears to be
nonlinear. The exponential fit provides a lifetime of 98 hr if 
the minimum area of the fragment is assumed to be 2 Mm$^2$ (the
uncertainty of the area measurements).  Observations show that the
fragment survived for at least 76~hr after fragmentation, as traces of
it were seen until the second half of July 3.

Figure~\ref{deca_flux} shows the temporal evolution of
the longitudinal magnetic flux of the parent sunspot and its fragment. 
Using the inversion results, this quantity is computed as 
$\phi$ = $\sum_i f_i \, B_i \, \cos \gamma_i \, S_i$, where $f_i$ 
represents the magnetic filling factor, $B_i$ and $\gamma_i$ the 
magnetic field strength and inclination, and $S_i$ the area of 
pixel $i$. The summation extends over all pixels. The magnetic flux 
of the parent remains nearly a constant at -5.0$\times$10$^{21}$~Mx
over a duration of 61 hr up to the point of fragmentation. 
The flux is then observed to decay strongly at a rate of -4.9$\times$10$^{15}$~Mx~s$^{-1}$
till the disappearance of the fragment.
The above value is consistent with that obtained by \citet{Deng2007} 
and \citet{Kubo2008}. After the disappearance of the fragment, the 
decay rate increases to -7.3$\times$ 10$^{15}$~Mx~s$^{-1}$.

The magnetic flux of the fragment (bottom panel 
of Figure~\ref{deca_flux}) amounts to -6.3$\times$10$^{20}$~Mx
at the time of separation. By July 2 it decreased down to 
-2.2$\times$10$^{20}$~Mx. The decay of the penumbra in the fragment could
be attributed to the order of magnitude deficit in the magnetic flux.

\subsection{Rotation of Fragment about Parent Sunspot}
\label{sunspot_rot}
During the evolution of the sunspot, the region enclosed by the light
bridge appears to rotate in an anti-clockwise direction with respect
to the main umbral core, as is evident from Figure~\ref{sunspot}. The
rotation described in this section refers to the orbiting motion of
the fragment around the center of the parent sunspot
\citep{Yan2008}. In order to determine the rotation of the fragment,
two points connecting the central axis and the rotating periphery need
to be located.  The centroid of each umbral core is calculated as
\begin{eqnarray}
x_c &=& \frac{\sum_{i,j} x_{i,j}I(i,j)}{\sum_{i,j}I(i,j)},\\
y_c &=& \frac{\sum_{i,j} y_{i,j}I(i,j)}{\sum_{i,j}I(i,j)}.
\end{eqnarray}

\begin{figure}[!h]
\centerline{
\includegraphics[angle=180,width = 0.5\textwidth]{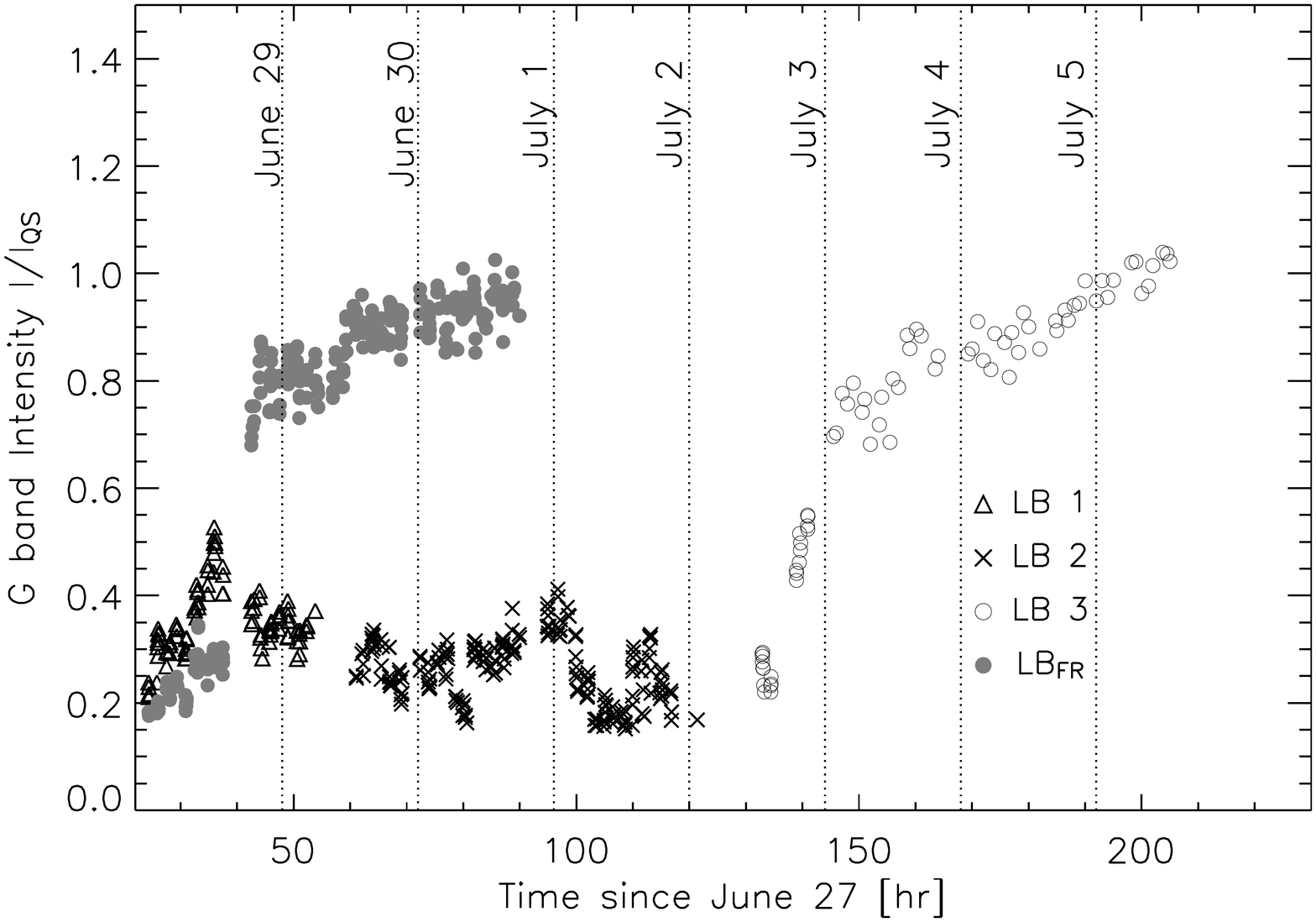}
}
\vspace{-10pt}
\caption{Temporal evolution of the mean G-band intensity of all 
sunspot LBs observed in NOAA AR 10961. }
\label{gban}
\end{figure}

The rotation angle is measured counter-clockwise from solar South
(vertically downwards) to the line joining the two centroids. 
Over the course of 5 days the fragment rotated about the parent by nearly
$60^\circ$ with an average rotation speed of $13^\circ$/day. This is consistent with
that obtained by \citet{Brown2003} for AR 9280\footnote{Incidentally, 
AR 9280 elongated and separated into two pores during its transit.}.
The separation between the parent and fragment, estimated as the
distance between their centroid positions, increased
almost linearly from approximately 10\arcsec\/ to 23\arcsec\/ in 
5 days. As the rotation of the fragment is more pronounced than its parent, 
it would suggest that the sunspot as a whole is comprised of several 
individual magnetic strands rooted to a common flux system rooted deeper in the 
photosphere \citep{Gonzalez2012}.

\subsection{Role of other LBs in Sunspot Evolution}
\label{lbs}
In addition to LB$_{\textrm{\tiny{FR}}}$, there were several other LBs
that were observed during the sunspot's lifetime but did not cause
fragmentation. This section compares various physical properties of
the LBs seen in the active region to ascertain why fragmentation occurred.
Figure~\ref{gban} shows the time variation of G-band
intensity in LB1, LB2 and LB3, the three LBs described in
Section~\ref{evol} and marked in Figure~\ref{sunspot}. To allow
comparisons, the curve corresponding to LB$_{\textrm{\tiny{FR}}}$ is
also included. The G-band intensities of LB1 and LB2 remain within
0.53$I_{\textrm{\tiny{QS}}}$ and do not exhibit any specific trend as
a function of time. By comparison, the light curve of LB3 appears very
similar to that of LB$_{\textrm{\tiny{FR}}}$, with the intensity
rising from umbral to near photospheric values in about 35 hr. The
photometric properties of both LB3 and LB$_{\textrm{\tiny{FR}}}$ are
consistent with each other as indicated in Table~\ref{tab_lb}.

\begin{table}[!h]
\begin{center}
\caption{Physical properties of sunspot LBs in NOAA AR 10961}
\label{tab_lb}
\begin{tabular}{llrrrr}
\hline
\multicolumn{2}{l}{Parameter}      & LB1      & LB2    & LB3       & LB$_{\textrm{\tiny{FR}}}$ \\
\hline
                                   & Min   &  0.21    &  0.15  & 0.22      &  0.18     \\ 
G-band intensity [$I_{\textrm{\tiny{QS}}}$] & Max   &  0.53    &  0.41  & 1.09      &  1.02     \\
				   & Mean  &  0.35    &  0.26  & 0.75      &  0.73     \\  
\hline
                                   & Min  &  1160    & 1390   & 680      & 630       \\
$B$ [G]                            & Max  &  2150    & 2130   & 2330     & 1880      \\   
                           	   & Mean  &  1720    & 1800   & 1620     & 1280      \\
        
\hline
                                   & Min   &  139     & 148  & 108    & 110     \\
$\gamma$ [deg]      		   & Max   &  172     & 168  & 166    & 176     \\
                                   & Mean  &  158     & 159  & 135    & 150     \\
\hline
Lifetime [hr]        &  & 30       & 60     & $>$80    &  45       \\
Area [Mm$^2$]        &  & 7.3      & 6.3    & 11.1       & 8.6$^\ast$        \\
Umbral Fraction [\%] &  & 23.7      & 15.6    & 38.4   & 5.9        \\
\hline
\end{tabular}
\end{center}
$^\ast$Area estimated on 18:31 UT June 28, 2007 after LB was formed.
\end{table}

The average field strengths in LB1 and LB2 exceed 1.5 kG, with a
minimum field strength of more than 1.1 kG. In LB3, the corresponding values
are 1.6 and 0.7 kG, respectively. This is comparable to
LB$_{\textrm{\tiny{FR}}}$, whose average and minimum field strengths
are 1.3 and 0.6 kG, 34~hr prior to fragmentation. In addition, the
field in LB1 and LB2 is mostly vertical with a mean inclination
of about 160$^\circ$.
This can be attributed to the fact that LB1 and LB2 morphologically
resemble umbral dots and penumbral filament intrusions.

The LB lifetimes range from 1 to 2.5 days. LB3 in particular is quite
long lived, since it was observed to be intact during the sunspot's
transit across the Western limb on July 7, 2007, implying a duration
of more than 80 hr. The lifetime of LB$_{\textrm{\tiny{FR}}}$ is
estimated to be $\sim$45 hr, from the moment it isolated the umbral
core until the sunspot fragmentation.  The area of the LBs in NOAA AR
10961 varied between 6.3 and 8.6 Mm$^2$.  While LB3 had an area of 7
Mm$^2$ on July 3 at 3:03 UT, this increased significantly to 11 Mm$^2$
nearly 8 hr later as the LB developed an additional arm connecting to
the penumbra at its southern end. The fraction of the umbra isolated
by the LBs is shown in the last row of Table~\ref{tab_lb}. The area
bounded by LB1 and LB2 represents 24\% and 16\% of the total umbral
area, while LB3 divided the umbra nearly in half. By comparison,
LB$_{\textrm{\tiny{FR}}}$ segregated only 6\% of the umbra. Although 
LB3 had physical characteristics similar to LB$_{\textrm{\tiny{FR}}}$, it
did not fragment the spot.
 
\subsection{Evolution of Horizontal Motions}
\label{flow}
In this section we describe the horizontal proper motions of intensity 
features, in and around the sunspot, and how they were affected by 
the fragmentation process. The horizontal motions were determined 
using local correlation tracking \citep[LCT;][]{November1986,November1988,Fisher2008,Welsch2004},
a technique which computes the relative displacement of small sub-regions centered on a
particular pixel with sub-pixel accuracy using cross-correlation
techniques. The sub-regions are apodized by a Gaussian window whose
full-width at half-maximum is roughly the size of the structures that
need to be tracked. Knowing the displacement and the time interval,
the horizontal speed for each pixel can be determined.  First, the
aligned G-band data sets are filtered for acoustic waves using a phase
velocity cut-off value of 6 km/s
\citep{Title1989}. The filtered images are then subject to the tracking
routine. After experimentation, an apodizing window of 1\arcsec\/
width and a time difference of 2 minutes between two images were
chosen. To reduce the noise in the measurements, 5 velocity images
were averaged. The LCT input parameters are similar to the ones used
by \citet{Santiago2010} and \citet{Meetu2011} on {\em Hinode} G-band
filtergrams.

\begin{figure*}[!ht]
\centerline{
\hspace{30pt}
\includegraphics[angle=180,width = 0.87\textwidth]{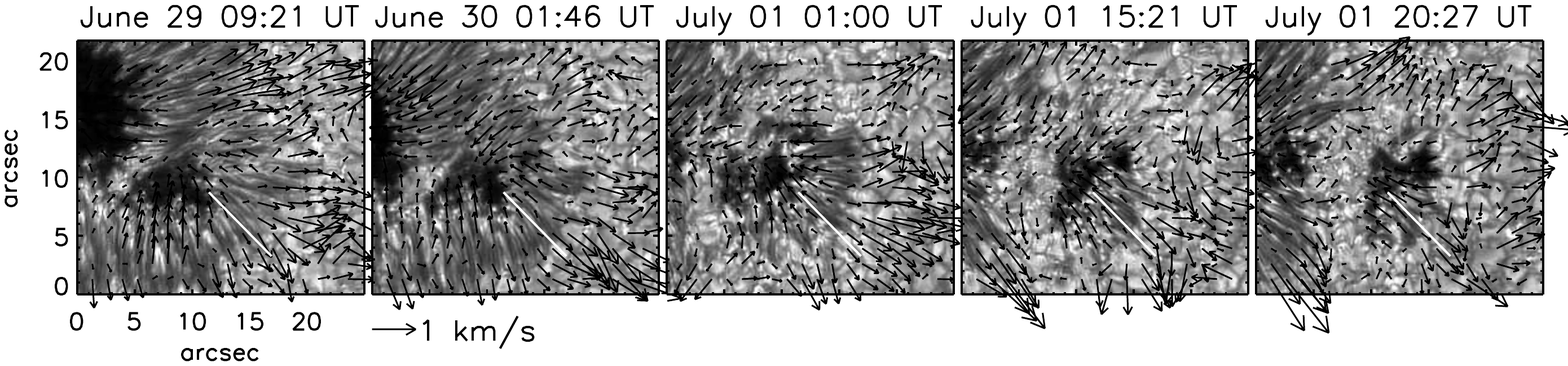}
}
\vspace{-293pt}
\centerline{
\hspace{30pt}
\includegraphics[angle=180,width = 0.87\textwidth]{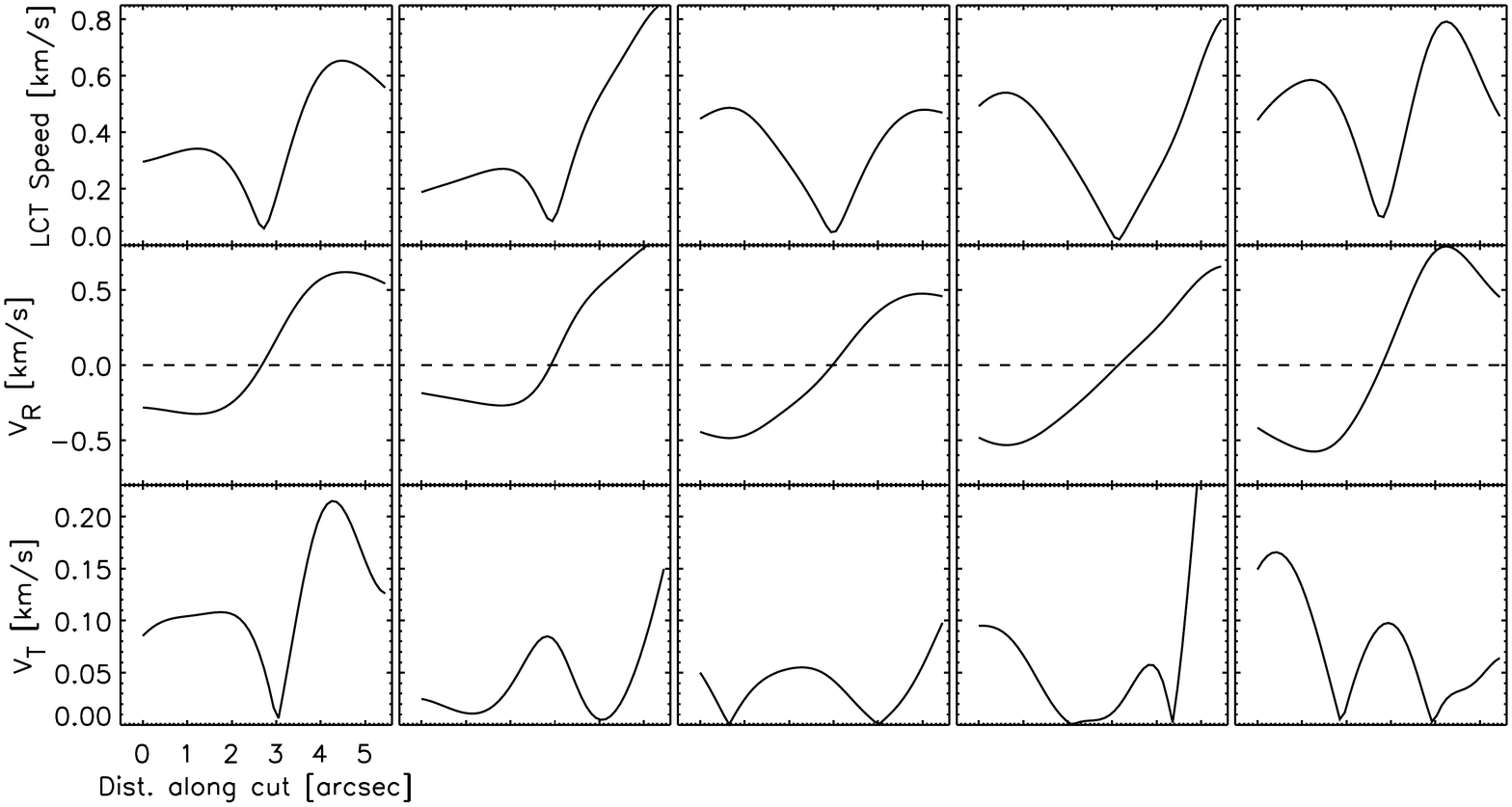}
}
\caption{Continuation of the horizontal motions in the decaying fragment 
after separation from the parent spot. {\em First row:} Horizontal
motion maps on different days overlaid on the G-band images. Arrows
have been drawn only for pixels where the speed is greater than 70
m~s$^{-1}$. The {\em white} line represents the cut along which the
horizontal speed and the radial ($V_R$) and tangential ($V_T$)
components of the velocity vector are shown in the second, third and
fourth rows, respectively.}
\label{lct_cont}
\end{figure*}

Horizontal proper motions in a sunspot are characterized by the presence of 
inward and outward directed flows originating from a region of 
divergence, or dividing line \citep[DL;][]{Sobotka1999,Sobotka2001}, 
located at 0.6--0.7 penumbral radii. Since these motions
change significantly in the region of fragmentation, remaining
fairly uniform in the rest of the AR, we restrict our
analysis of the horizontal motions in and around the fragment 
in the following sections.

\subsection{Continuance of Horizontal Motions in Fragment}
\label{conti}
As the width of LB$_{\textrm{\tiny{FR}}}$ increases on June 30 and the
separation between the sunspot and its fragment widens, the average
speeds of proper motions in the intermediate granulation region are
seen to be less than 100 m~s$^{-1}$. 
Radial motions are seen in the penumbra of the fragment even after separating
from the parent. Figure~\ref{lct_cont} shows horizontal
velocity maps at different stages of evolution of the fragment. The
horizontal velocity vectors have been overlaid on the G-band images in
the top panel. Below each G-band snapshot we plot the variation of the
horizontal speed, as well as the radial and tangential components of
the velocity along a radial cut in the penumbra (white line). The
radial component is measured with respect to the cut and is negative
and positive for inward and outward motions,
respectively. Figure~\ref{lct_cont} illustrates the following: i) the
inward and outward motions from the DL is present even after
fragmentation; ii) such a radial pattern is only observed in penumbral
filaments wherever present in the decaying fragment; and iii) inward
motions of 100--150m~s$^{-1}$ are seen at the location of
fragmentation which also lacks a penumbra. These inward motions are
weaker than those typically seen in the inner penumbra of the sunspot.

\begin{figure}[!h]
\centerline{
\includegraphics[angle=180,width = 0.5\textwidth]{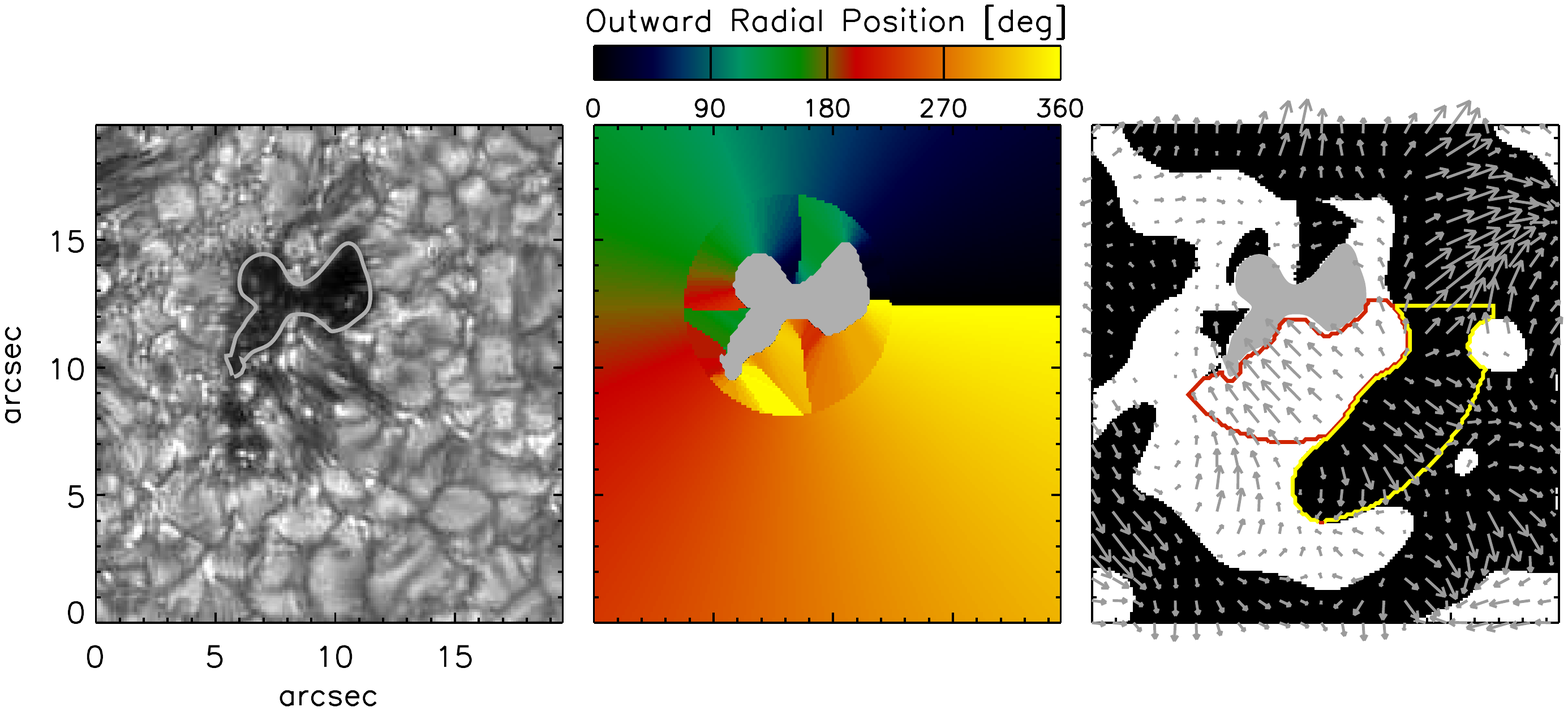}
}
\vspace{-65pt}
\caption{Horizontal motions around fragment with final traces of 
penumbra. {\em Left panel:} G-band image of fragment on July 1 at
20:39 UT. The grey contours represent the umbra-penumbra boundary of
the fragment. {\em Middle panel:} Azimuth corresponding to the
position vector measured in the outward direction from fragment
centroid or umbra-penumbra boundary. The azimuth increases in the
anti-clockwise direction with $0^\circ$ and $180^\circ$ pointing
towards solar West and East as shown in the color bar. {\em Right
panel:} Binary mask representing inward ({\em white}) and outward
({\em black}) radial motions surrounding the fragment. The horizontal
velocity vectors are indicated with grey arrows. The thick red/yellow
contours span the azimuth range from 220 to $360^{\circ}$ and
represent the inflow/outflow regions.}
\label{pore_penumbra}
\end{figure}

\begin{figure}[!h]
\centerline{
\hspace{15pt}
\includegraphics[angle=180,width = 0.475\textwidth]{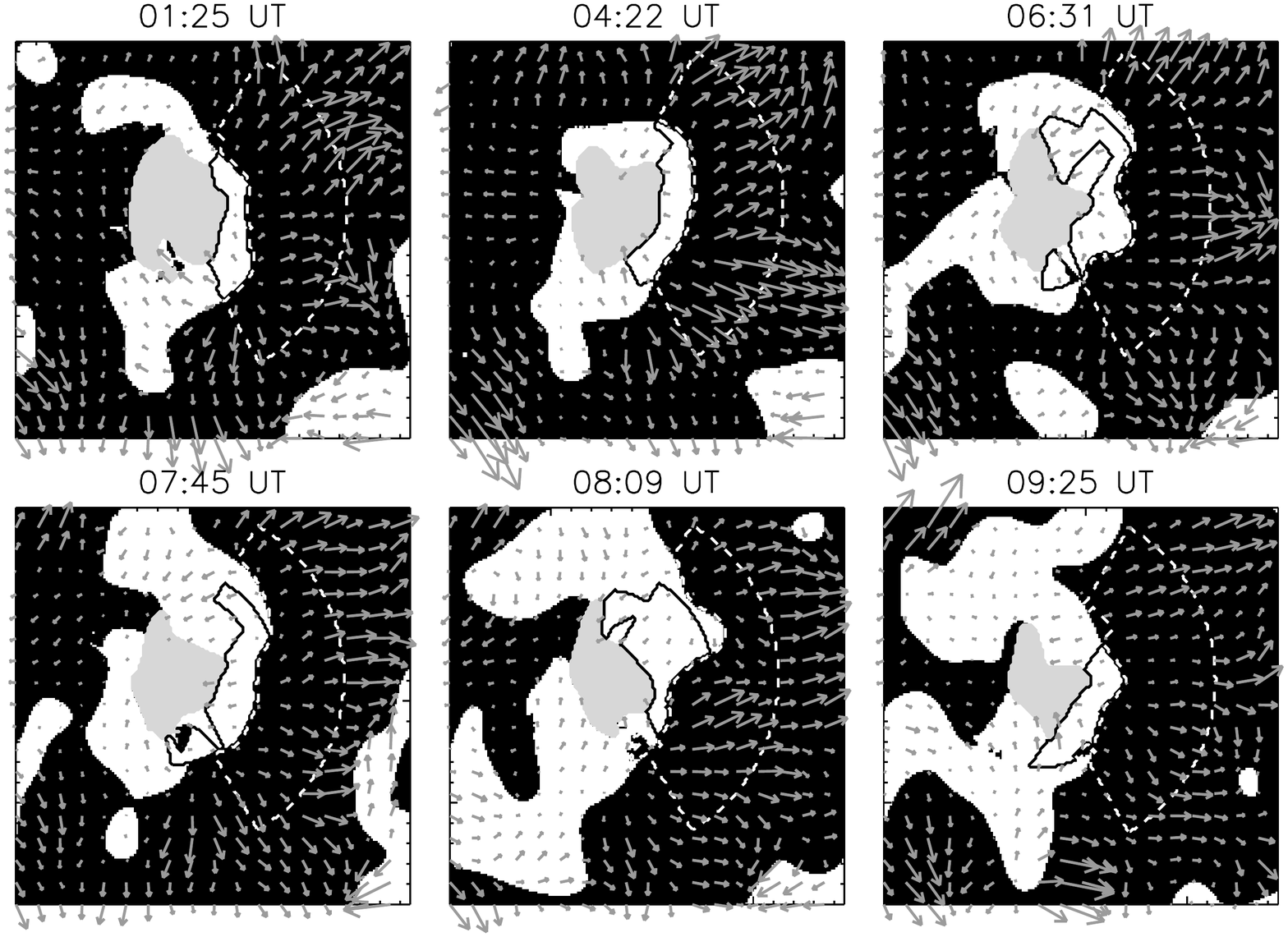}
}
\vspace{-10pt}
\centerline{
\hspace{15pt}
\includegraphics[angle=180,width = 0.475\textwidth]{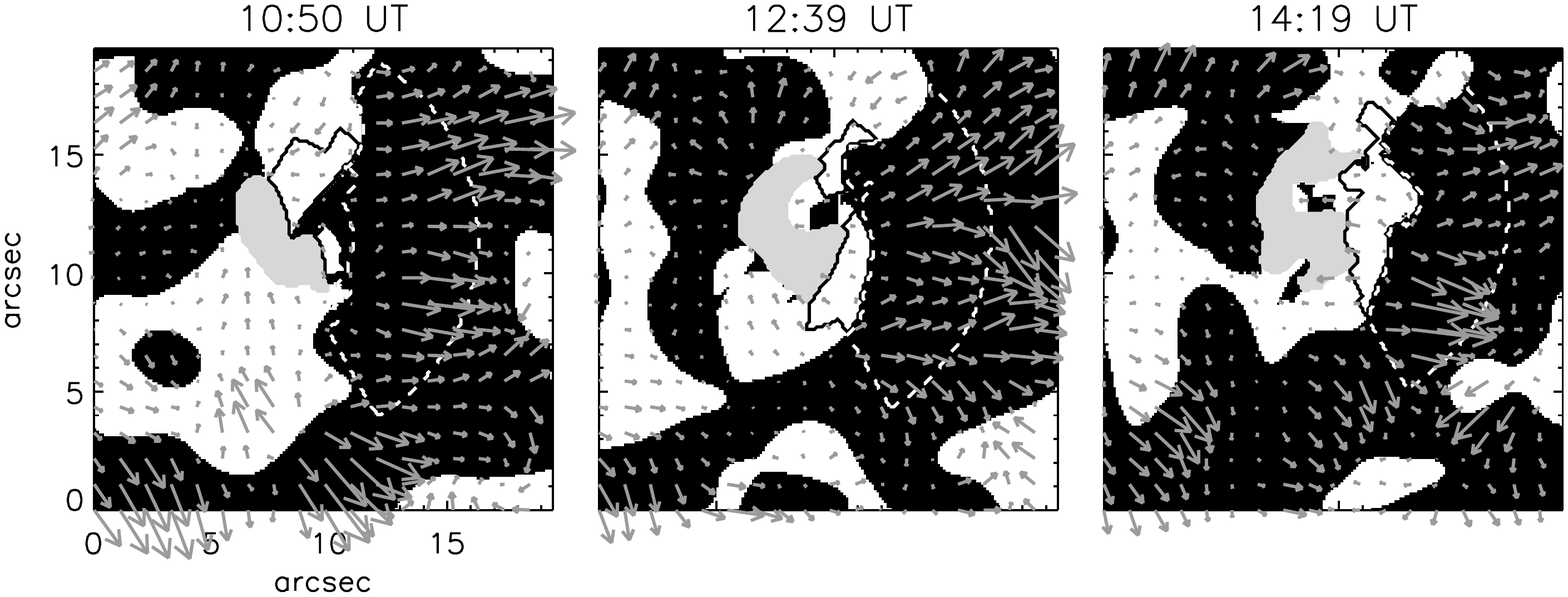}
}
\vspace{-85pt}
\caption{Binary image representing inward and outward radial motions 
surrounding the pore on July 2.  The horizontal velocity vectors are
indicated with grey arrows except for the pore which has been shaded
in light grey. The black contour outlined on the western interface of
the pore corresponds to the inflow region chosen for analysis. The
contour has a $\pm 60^\circ$ azimuthal span centered at $0^\circ$
azimuth.}
\label{pore_binary}
\end{figure}

\subsubsection{Horizontal Motions during the Pore Phase}
\label{pore}
Here we compare the nature of the horizontal motions in the pore on
July 2 and in the fragment comprising a penumbra a day earlier.  In
order to isolate regions of inward and outward motions surrounding the
pore, the angle $\beta$ between the position vector and the velocity
vector at each pixel was determined. The former is directed from the
pore centroid or the pore-QS boundary outwards. The azimuth of the
position vector is the angle it makes with solar West, increasing in
the counter-clockwise direction. For radial distances greater than
4\arcsec\/, the azimuth is calculated using the pixel location and the
pore's centroid. For pixels closer to the pore, the azimuth is
assigned the value of the point on the pore-QS boundary lying nearest
to the pixel of interest. The above procedure is similar to the one adopted by
\citet{Santiago2010}. Once the position angle is known, $\beta$ is
calculated such that it is 180$^\circ$ and $0^\circ$ for pure radial
and outward motions, respectively. Binary maps are then obtained
assigning 1 for $90^\circ<\beta<180^\circ$ and 0 for
$0^\circ<\beta<90^\circ$. The middle panel of
Figure~\ref{pore_penumbra} shows the azimuths of the position vector
on June 1 at 20:39 UT, while the right panel displays the binary image
depicting inward and outward motions. The red and yellow contours span
the azimuth range from 220$^{\circ}$ to 360$^{\circ}$.

\begin{figure}[!h]
\centerline{
\hspace{35pt}
\includegraphics[angle=0,width = 0.5\textwidth]{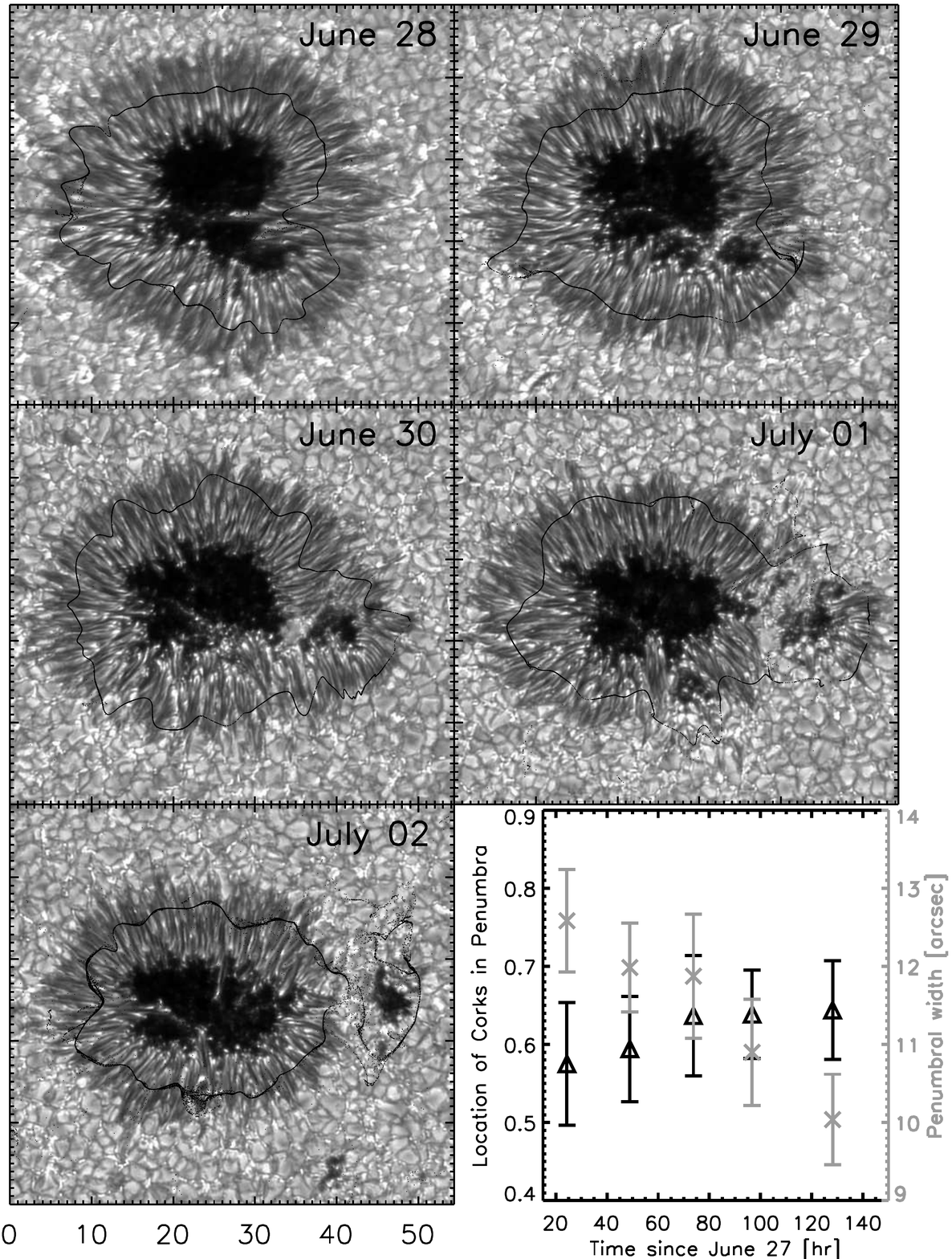}
}
\vspace{5pt}
\caption{Cork maps depicting the evolution of the sunspot and its horizontal 
flow field. Each cork map was constructed using 25,600 artificial tracer 
particles, shown in black dots, that were tracked backward in time for 
24 hr. {\bf{Bottom right panel: }} Evolution of the position of the 
DL ({\em black tirangles}) in terms of the normalized penumbral distance and 
penumbral width ({\em grey crosses}). The vertical bars correspond to the $rms$ value.}
\label{cork}
\end{figure} 

Figure~\ref{pore_binary} shows a sequence of binary maps on July 2,
when the fragment was reduced to a pore.  The pore has been masked in
light grey and the grey arrows represent the velocity vectors. The
figure clearly shows a region of inward motions surrounding the pore
on the western side.  The inward motions are interrupted by slower
outward motions on the eastern side, possibly due to the presence of
the larger parent sunspot. The black contour represents a region of
inward motions having a $\pm 60^\circ$-span in azimuth around the
western edge of the pore. This is the region selected for further
analysis.  Since the binary maps on July 2 do not show clear
differences with the ones on July 1 when the penumbra was still
present, we compare the magnitude of the inward and outward motions
during the pore phase with that of the previous day when the penumbra
was intact.

In the pore, the mean speed of the inward motions is 
0.1--0.3~km~s$^{-1}$, while the fragment has $\sim$0.4~km~s$^{-1}$ on
July 1. The radial distance of the DL from the pore boundary is
$\sim$1\farcs6. In the case of the fragment, the DL lies further out
at 3\arcsec. The DL is located at even larger radial distances in the
parent sunspot that has a well formed penumbra, up to 6\arcsec\/ from
the umbra. These results are in good agreement with the findings of
\citet{Deng2007} and \citet{Santiago2010}.  The differences between 
the horizontal motions surrounding the fragment when it is a pore and
when it shows a penumbra are: i) the speed of the inward motions,
which is systematically smaller by $\sim$0.25 km~s$^{-1}$ in the pore
as revealed by the LCT maps on July 2, and ii) the location of the DL
separating the inward and outward motions, which lies much closer to
the pore boundary in the former.

\subsubsection{Persistent Divergence Locations}
\label{cor}
In the previous section, the inward and outward motions surrounding
the pore were compared with those of the fragment on the previous
day. A common feature that was observed in both cases is the region of
divergence separating the two types of motions, although their nature
and properties are distinct from one another. In order to identify the
persistent locations of divergence as the sunspot evolved we use the
method of corks described below.

We start out with 25,600 artificial trace particles scattered
uniformly and allowed to advect in the flow field. These corks are
traced backwards in time for 24 hr. As a consequence of
the long time duration, the corks appear to collect at the strongest
centers of divergence which in this case lie in the outer penumbra of
the sunspot. The resulting locations of divergent motion are shown 
in Figure~\ref{cork}. Furthermore, the shape of the DL gets stretched 
in the direction of separation as the fragmentation commences. Note 
that the DL remains fairly constant in the rest of the sunspot as 
described earlier. The regions where the DL appears discontinuous 
coincide with places which lack a penumbra or where photospheric 
granulation intervenes. This is seen particularly on July 1 near the 
southern region of the parent sunspot and the anchorage points of the 
light bridge where the fragment separated from the spot. The existence 
of a DL in the penumbra of the fragment even after the break up process 
is consistent with the observations shown in Figure~\ref{lct_cont}. On 
July 2, the DL closed in around the parent spot, hinting at the 
re-establishment of the penumbra. The DL was also present in the pore 
on July 2, but at a smaller distance from the pore-QS boundary than in 
the sunspot.

The bottom right panel of the figure indicates that as the sunspot 
evolves, the DL has a weak tendency to move outwards. The position 
of the DL in terms of the normalized penumbral distance, varies 
from 0.58$\pm$0.16 to 0.64$\pm$0.13 from June 28 to July 2 respectively. 
There is also a systematic reduction in the width of the penumbra
as the sunspot decays, decreasing from 12.6$\pm$1.3 arcsec to 
10.0$\pm$1.2 arcsec over a course of 5 days. Those pixels that lie 
outside the sunspot or close to the fragment umbral core were not 
considered in the above calculation.

\section{Discussion}
\label{summary}
We followed the evolution of a sunspot in NOAA AR 10961 for 8 days in June/July 
2007 using high resolution imaging and spectropolarimetric observations from {\em
Hinode}. The observations show the formation of a LB that fragmented
the sunspot in two parts. The LB formed as a consequence 
of penumbral filament intrusion into the umbra. The fragmentation process 
began with the depletion of the penumbra at the anchorage points of the
light bridge. Consequently, the characteristic pattern of proper
motions in and around sunspots was also disrupted. The presence of 
the fragment inhibited its restoration.  This could be due to the
inability of magnetic fields to rise to the surface by buoyancy
because of the intervening convective wall that extended beneath the
photosphere between the parent sunspot and the fragment.

The precursors to penumbral formation are elongated granules that
develop structures resembling penumbral filaments close to the edge of
the sunspot boundary, as observed by \citet{Schlichenmaier2010,
Schlichenmaier2010b} and \citet{Lim2011}. These features could be tiny
$\Omega$-shaped magnetic loops. In addition, one finds the 
formation of the penumbra in the sunspot to be closely related to 
the magentic field inclination at the penumbra-QS boundary 
as seen in the post fragmentation phase.

So does the presence of a light bridge herald the fragmentation of a sunspot
and what are the conditions for this to happen? The significant 
field strength reduction in the LB along with the 
presence of granulation is suggestive of strong convection which might 
have triggered the expulsion and fragmentation of the smaller umbral
core. \citet{Katsukawa2007} point out that the process of LB
formation, fragments the sunspot by injecting hot, weakly magnetized
gas into the umbra. However, the LB analyzed by \citet{Katsukawa2007}
remained an extension of the penumbra and did not separate 
individual umbral cores. Furthermore, the intensity and field strength did
not evolve to QS values and the sunspot remained a single coherent structure
during its transit. This does not imply however, that LBs which have matured 
into photospheric structures will in turn fragment the sunspot. LB3, which 
was seen in the sunspot after the fragmentation, had similar characteristics to
LB$_{\textrm{\tiny{FR}}}$ and in this regard it ought to have split
the sunspot within a time scale of 3-4 days after it had reached QS
conditions, but it did not. The failure of fragmentation might be
attributed to the fraction of the umbra that LBs isolate in
general. This could decide if individual fragments can remain stable
to convective motions. In our case, we cannot rule out the possibility
that the sunspot could have fragmented soon after its passage across
the Western limb. Although the presence of photospheric conditions in
a light bridge is a necessary condition for fragmentation, it is not a
sufficient one. 

The fragmentation of sunspots, seen as a transition of their
sub-structures, demonstrates a strong interplay between convection and
magnetic fields extending over a large range of spatial and temporal
scales. The next step would be to carry out a similar investigation on
the formation of sunspots, which would be crucial to understand the
organization and stability of magnetic fields at the solar surface.

\acknowledgments
Our sincere thanks to the {\em Hinode} team for providing the data
used in this paper. Hinode is a Japanese mission developed and
launched by ISAS/JAXA, with NAOJ as domestic partner and NASA and STFC
(UK) as international partners. It is operated by these agencies in
co-operation with ESA and NSC (Norway). Hinode SOT/SP Inversions were
conducted at NCAR under the framework of the Community
Spectro-polarimtetric Analysis Center. We thank the anonymous referee 
for useful comments and suggestions. R.E.L. is grateful for 
the financial support from the German Science Foundation (DFG) under 
grant DE 9422. Our work has been partially funded by the Spanish 
MICINN through projects AYA2009-14105-C06-06 and by 
Junta de Andaluc\'{\i}a through project P07-TEP-2687.

\end{document}